\documentclass[a4paper,fleqn]{article}

\usepackage[utf8]{inputenc}
\usepackage[english]{babel}

\usepackage{graphicx}
\graphicspath{{fig/}{../fig/}}

\usepackage{csquotes}
\usepackage{xspace}

\usepackage{amsfonts,amsmath,amssymb}
\usepackage[usenames,dvipsnames]{xcolor}

\usepackage[space]{grffile}
\usepackage{latexsym}
\usepackage{textcomp}
\usepackage{longtable}
\usepackage{tabulary}
\usepackage{booktabs,array,multirow}

\usepackage[round]{natbib}
\usepackage{url}

\usepackage{siunitx}
\usepackage{placeins}
\usepackage{subcaption}
\usepackage{subfiles}
\usepackage{wrapfig}
\usepackage{diagbox}
\usepackage{todonotes}

\DeclareMathOperator\sinc{sinc}
\renewcommand{\d}{\mathrm{d}}
\newcommand{\e}{\mathrm{e}}

\definecolor{darkred}{RGB}{150,0,0}
\definecolor{darkblue}{RGB}{0,0,100}
\definecolor{darkgreen}{RGB}{0,80,0}

\usepackage[top=1in, bottom=1.25in, left=1.25in, right=1.25in]{geometry}

\newcommand{\beginsupplement}{%
	\setcounter{table}{0}
	\renewcommand{\thetable}{S\arabic{table}}%
	\setcounter{figure}{0}
	\renewcommand{\thefigure}{S\arabic{figure}}%
}

\begin{document}
	
	\title{Revisiting the Identification of Wintertime Atmospheric Circulation Regimes in the Euro-Atlantic Sector}
	
	\author{Swinda K.J. Falkena$^1$, Jana de Wiljes$^{1,2}$, Antje Weisheimer$^{3,4}$, \\ Theodore G. Shepherd$^5$}
	\date{}
	
	\maketitle
	
\noindent{$^{1}$Department of Mathematics and Statistics, University of Reading, Reading, UK\\
		$^{2}$Institute for Mathematics, University of Potsdam, Potsdam, Germany\\
		$^{3}$European Centre for Medium-Range Weather Forecasts (ECMWF), Reading, UK\\
		$^{4}$National Centre for Atmospheric Science (NCAS), University of Oxford, Department of Physics, Atmospheric, Oceanic and Planetary Physics (AOPP), Oxford, UK\\
		$^{5}$Department of Meteorology, University of Reading, Reading, UK\\
	}
	
	{\centering
		\textbf{Corresponding author}: {s.k.j.falkena@pgr.reading.ac.uk}\\[10mm]
	}

\begin{abstract}
Atmospheric circulation is often clustered in so-called circulation regimes, which are persistent and recurrent patterns. For the Euro-Atlantic sector in winter, most studies identify four regimes: the Atlantic Ridge, the Scandinavian Blocking and the two phases of the North Atlantic Oscillation. These results are obtained by applying $k$-means clustering to the first several empirical orthogonal functions (EOFs) of geopotential height data. Studying the observed circulation in reanalysis data, it is found that when the full field data is used for the $k$-means cluster analysis instead of the EOFs, the optimal number of clusters is no longer four but six. The two extra regimes that are found are the opposites of the Atlantic Ridge and Scandinavian Blocking, meaning they have a low-pressure area roughly where the original regimes have a high-pressure area. This introduces an appealing symmetry in the clustering result. Incorporating a weak persistence constraint in the clustering procedure is found to lead to a longer duration of regimes, extending beyond the synoptic timescale, without changing their occurrence rates. This is in contrast to the commonly-used application of a time-filter to the data before the clustering is executed, which, while increasing the persistence, changes the occurrence rates of the regimes. We conclude that applying a persistence constraint within the clustering procedure is a superior way of stabilizing the clustering results than low-pass filtering the data.

\end{abstract}%

\section{Introduction}
\label{sec:intro}
 
 The study of atmospheric circulation, or weather, regimes has a long history. Starting from the 1940s, when the German weather service developed a set of weather types classifying the daily synoptic circulation \cite[][]{DeutscherWetterdienst}, the concept of weather regimes as an expression of the low-frequency variability of the atmospheric circulation has been a topic of research. The underlying concept is that the weather itself is a stochastic process, whose statistics are strongly conditioned on the weather regime. From around 1990 onwards different clustering methods have been used to identify these persistent and recurrent circulation patterns \cite[e.g.][]{Mo1987,Vautard1990, Molteni1990}, primarily focussing on the wintertime Northern Hemisphere. Later specific sectors of the Northern Hemisphere, primarily the Euro-Atlantic sector \cite[e.g.][]{Michelangeli1995, Kageyama1999} and the Pacific-North American sector \cite[e.g.][]{Straus2007, Riddle2013, Amini2018}, as well as the Southern Hemisphere \cite[e.g.][]{Mo2000} have been studied, along with the relation of circulation regimes with e.g. climate change \cite[][]{Corti1999} and regional weather \cite[][]{Cassou2005}. Also more limited areas have been considered \cite[e.g.][]{Robertson1999}.

Initially different clustering methods, such as hierarchical clustering \cite[e.g.][]{Cheng1993}, using the analysis of the probability density function \cite[e.g.][]{Kimoto1993b}, or $k$-means clustering \cite[e.g.][]{Michelangeli1995}, were used to identify the atmospheric circulation regimes. The last method, $k$-means clustering, has subsequently become the most used approach for identifying the regimes in atmospheric data \cite[][]{Hannachi2017}. The number of clusters $k$ has to be set a priori, making finding the optimal number of regimes part of the problem. Commonly-used methods to do this are the verification of significance by using synthetic datasets \cite[e.g][]{Dawson2012} or looking at the reproducibility and consistency when the algorithm is run multiple times \cite[e.g.][]{Michelangeli1995}. Nearly always the data is first projected onto the first several Empirical Orthogonal Functions (EOFs), after which the clustering algorithm is applied to the time series of these EOFs \cite[e.g.][]{Vautard1990, Ferranti2015}. In addition some studies apply a low-pass time filter to the data to remove the high frequency, noisy oscillations and focus on the low-frequency behaviour \cite[e.g.][]{Straus2007, Grams2017}. This method enforces a higher persistence of the regimes compared to standard $k$-means clustering which is independent of the time-ordering of the data.

Since clustering methods represent a projection of the data to a lower dimensional state space, applying clustering to the already filtered data of EOFs means a projection of the data is done twice. Thirty years ago, this approach was necessary because computational limitations did not allow using the full field dataset. However this is no longer a constraint. Nevertheless most studies continue to follow the original approach and use EOFs. As EOFs give the modes associated with the most variability, while clusters give the recurrent patterns, the means of dimension reduction is quite different. The question thus arises of what the effect of this double filtering is on the resulting atmospheric circulation regimes. Similarly, applying a low-pass filter to remove the high-frequency behaviour before the cluster analysis also means the data is filtered twice. This is likely to not only affect the persistence, but also the occurrence of the found regimes and possibly the clusters themselves, thus also raising the question of how strong this effect is.

In this paper we compare the results of $k$-means clustering using EOF data with the results found for the full field data, for the case of the wintertime Euro-Atlantic sector. We pay special attention to the optimal number of clusters $k$, as it has a large influence on the regimes that are found. Furthermore, we study the effect of time-filtering on the found regimes by comparing it with an adapted $k$-means clustering algorithm that incorporates a constraint on the regime duration to enforce persistence of the regimes. This novel algorithm, which is discussed in Section \ref{ssec:meth_clus}, does not change the data itself, but only the method to identify the regimes. Both comparisons give insight into the effect of filtering the data, either using EOFs or applying a low-pass filter, before applying the $k$-means clustering algorithm. We start by explaining the methods used, followed by a discussion of the differences and similarities of the results. In the end the findings are summarized and discussed.

\section{Methods}
\label{sec:meth}

We use 500 hPa geopotential height data from ERA interim on a 2.5$^\circ$ by 2.5$^\circ$ longitude-latitude grid for a domain covering the Euro-Atlantic sector, 20$^\circ$ to 80$^\circ$N and 90$^\circ$W to 30$^\circ$E \cite{Dee2011}. Daily data (00:00 UTC) is considered for the months December through March using 39 years of data (1979 - 2018). Deviations from a fixed background state are used throughout this period. The main argument for considering a fixed background state instead of a seasonally varying one is that when applying cluster analysis the data used is preferably as complete as possible to avoid any type of bias. This means that few to no assumptions, such as a seasonal cycle, are made in preparing the data to retain the information present in the data. Or, phrased differently, how can one compare two days if they are deviations with respect to a different background state? The risk of this approach is that seasonality affects the regimes that are found and introduces a bias in the occurrence and persistence. Thus there is a trade-off to be made between obtaining as large a sample size as possible whilst minimizing such effects. The rationale for the choice of the period December-March is based on the difference in background state between the different months and explained in the Supplementary Information. Based on this analysis, we do not expect the found regimes to be sensitive to the removal of the seasonal cycle. Differences in the occurrence and persistence of the regimes cannot be ruled out, and their behaviour throughout winter is discussed in the Supplementary Information.

\subsection{Clustering Methods}
\label{ssec:meth_clus}

The method used for the identification of circulation regimes in this study is $k$-means clustering using the standard Euclidian distance ($L_2$-norm) \cite[][]{Jain2010}. This method is applied to both the full field dataset as well as the time series of the first 5, 10, 15 and 20 EOFs. Furthermore, the method is also applied to the full field data after applying a 5- and a 10-day low-pass filter to remove high-frequency oscillations. The results for this time-filtered dataset are compared with those obtained by applying an adapted $k$-means algorithm to the unfiltered data that incorporates a persistence constraint in the clustering procedure itself. This persistent clustering method is described in what follows.

Given a dataset $\{x_t\}_{t\leq T}\in\mathbb{R}^m$, with $t$ time, $m$ the dimension of the data and $T$ the span of the data given in time, the aim of any clustering method is to find a set of $k$ cluster centres that accurately describe the dataset based on some measure. Let $\Theta = (\theta_1, ..., \theta_k)$ be the set of parameters describing the $k$ cluster centres in either EOF or grid point space. Here $\Theta$ represents the different circulation regimes for 500 hPa geopotential height anomaly data $\{x_t\}_{t\leq T}$. To assess how well the cluster centres represent the data, a model distance functional $g(x_t, \theta_i)$, giving the distance between a cluster centre and a data point, is required. We use the standard $L_2$-distance weighted by the cosine of latitude \cite[][]{Chung1999}. In addition we consider the affiliation vector $\Gamma = (\gamma_1(t), ..., \gamma_k(t))$, which indicates the weight of a certain cluster at some point in time. In practice $\gamma_i(t)$ is nearly always either zero or one, indicating to which cluster that point belongs. This is because a linear optimization problem always has an optimal solution on the boundary of the admissable set \cite[][]{Cottle2017}. For this reason the affiliation vector is in general not considered when $k$-means clustering is discussed. Here we do consider this vector because it allows for the incorporation of persistence in the clustering procedure.

The task of identifying the atmospheric circulation regimes best representing the data means one has to find the optimal parameters for the cluster centres $\Theta$ and the affiliations of the data $\Gamma$. This is done by minimizing the averaged clustering functional \cite[][]{Franzke2009}
\begin{equation}
\label{eq:optfunc}
\mathbf{L}(\Theta, \Gamma) = \sum_{t=0}^T \sum_{i=1}^k \gamma_i(t) g(x_t,\theta_i),
\end{equation}
subject to
\begin{equation}
\label{eq:optfuncconstr}
\sum_{i=1}^k \gamma_i(t)  = 1, \quad \forall t\in[0,T], \qquad \gamma_i(t) \geq 0, \quad \forall t\in[0,T], i=1,...,k.
\end{equation}
This is what the $k$-means procedure is doing implicitly, where $\gamma_i(t)$ is assumed to be zero or one. Finding the minimum of this functional minimizes the within-cluster variance as $\mathbf{L}$ is a measure of the distance between the cluster centres and the data points assigned to it. Because the within-cluster variance is minimized simultaneously for all clusters, the distance between data points assigned to different clusters becomes large. In other words; the between-cluster variance is maximized.

This clustering functional does not yet incorporate any persistence; an arbitrary reshuffling of the data leads to exactly the same result. To include persistence in the clustering method we add a constraint on $\Gamma$ that limits the number of transitions between regimes that is allowed \cite[][]{deWiljes2014}. This constraint on the number of transitions between regimes, or switches, that are permitted throughout the whole time-series is:
\begin{equation}
\label{eq:constrpers}
\sum_{i=1}^{k} \sum_{t=0}^{T-1} | \gamma_i(t+1) - \gamma_i(t) | \leq C, \quad i=1,...,k, 
\end{equation}
for some constant $C$. The value of $C$ gives twice the number of switches allowed (for a transition $\theta_1 \rightarrow \theta_2$ both the switch out of $\theta_1$ and the one into $\theta_2$ are counted), so e.g. an average cluster length of five days corresponds to $C = 2\times\#\text{days}/5 \approx 1900$. In Table \ref{tab:Cday} the average regime duration corresponding to several values of $C$ are given. Note that for a first-order Markov process, which is a relatively accurate assumption for the unconstrained regime dynamics, the average regime duration $T_{av}$ can be related to the e-folding time scale $T_e$ according to $T_{av}  = 1/(1-p)$ and $T_e = -1/\log(p)$, where $p$ is the self-transition probability (see SI for details). Hence $T_e$ is slightly shorter than $T_{av}$ (e.g. 4.8 days for an average duration of 5.3 days), and they approach each other as $p \rightarrow 1$. The rationale behind this constraint is that in a chaotic atmospheric circulation not every data point can be straightforwardly assigned to a cluster. Some data points can be in-between clusters or outliers, e.g. transitioning between clusters or extreme events. $K$-means clustering assigns these points to the nearest cluster (by distance), while it can be more sensible to assign them to the same cluster as their neighbours if the distance to that cluster is also quite small. This is exactly what the constraint in Equation \ref{eq:constrpers} is doing for reasonable values of $C$.

\begin{table}
	\caption{\label{tab:Cday}The value of $C$ with corresponding average regime duration.}
	\centering
	\small{
	\begin{tabular}{c |ccccccccc}
		$C$ & 600 & 800 & 1000 & 1200 & 1400 & 1600 & 1800 & 2000 & 2200 \\
		\hline
		Average Duration (days) & 15.8 & 11.8 & 9.5 & 7.9 & 6.8 & 5.9 & 5.3 & 4.7 & 4.3
	\end{tabular}
}
\end{table}

The minimization of the clustering functional $\mathbf{L}$ taking into account the persistence constraint is done in two steps which are iterated until convergence:
\begin{enumerate}
	\item For fixed $\Theta$, minimize $\mathbf{L}$ over all possible values of $\Gamma$
	\item For fixed $\Gamma$, minimize $\mathbf{L}$ over all possible values of $\Theta$
\end{enumerate}
The first part is done by linear programming using the Gurobi package for python \cite[][]{Gurobi}. The second part is done by $k$-means clustering. The computation is terminated when the difference between consecutive $\mathbf{L}$ becomes smaller than a set tolerance.

\begin{figure}[h]
	\centering
	\begin{subfigure}{.39\textwidth}
		\centering
		\includegraphics[width=1.\textwidth]{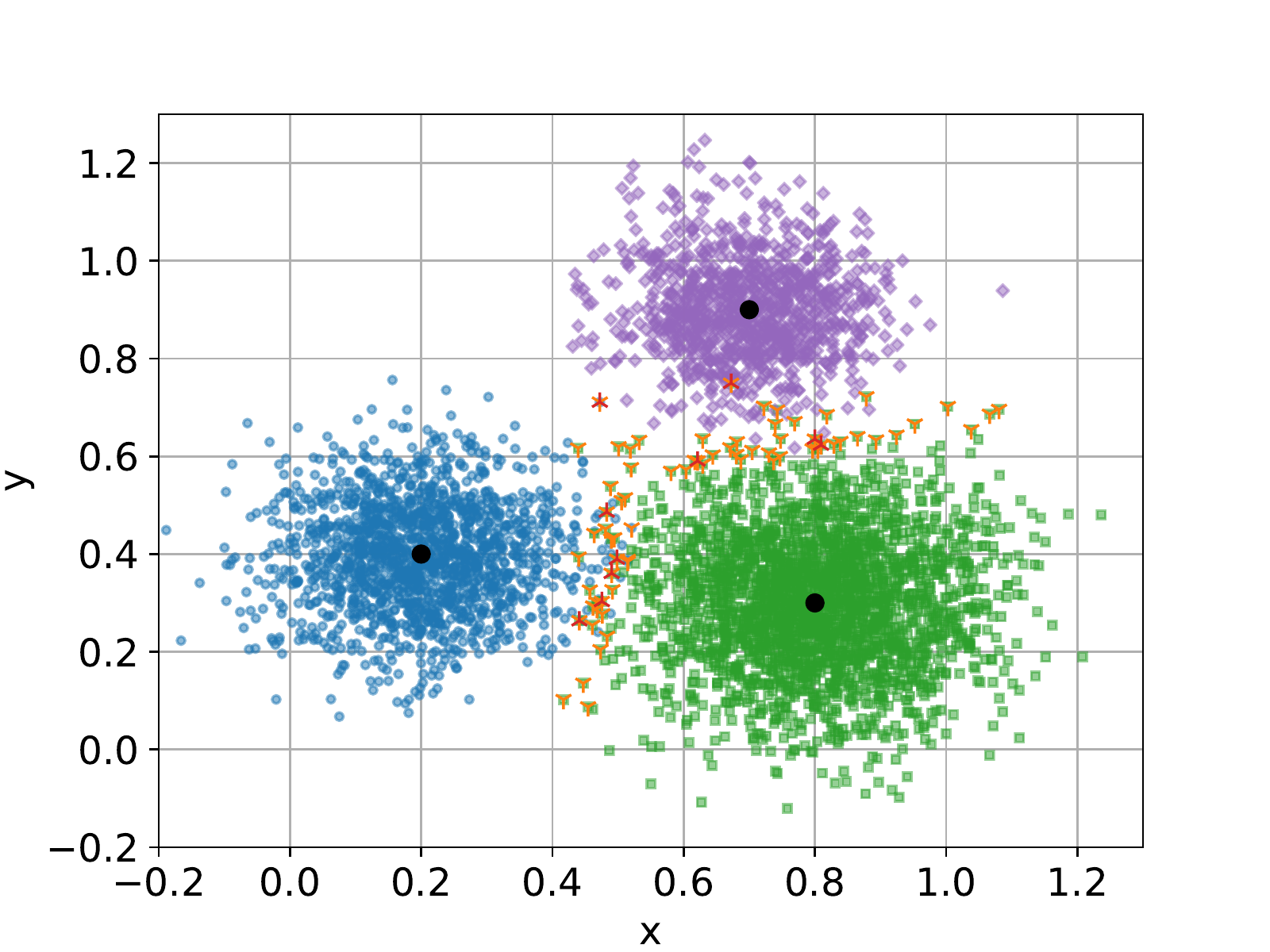}
		\caption{\label{fig:toymod}The data points wrongly assigned by standard $k$-means clustering are shown in orange (Y-shape) and the points that are still wrongly assigned when the persistence constraint is incorporated are shown in red (inverted Y-shape).}
	\end{subfigure}
\hspace{0.01mm}
	\begin{subfigure}{.59\textwidth}
		\centering
		\includegraphics[width=1.\textwidth]{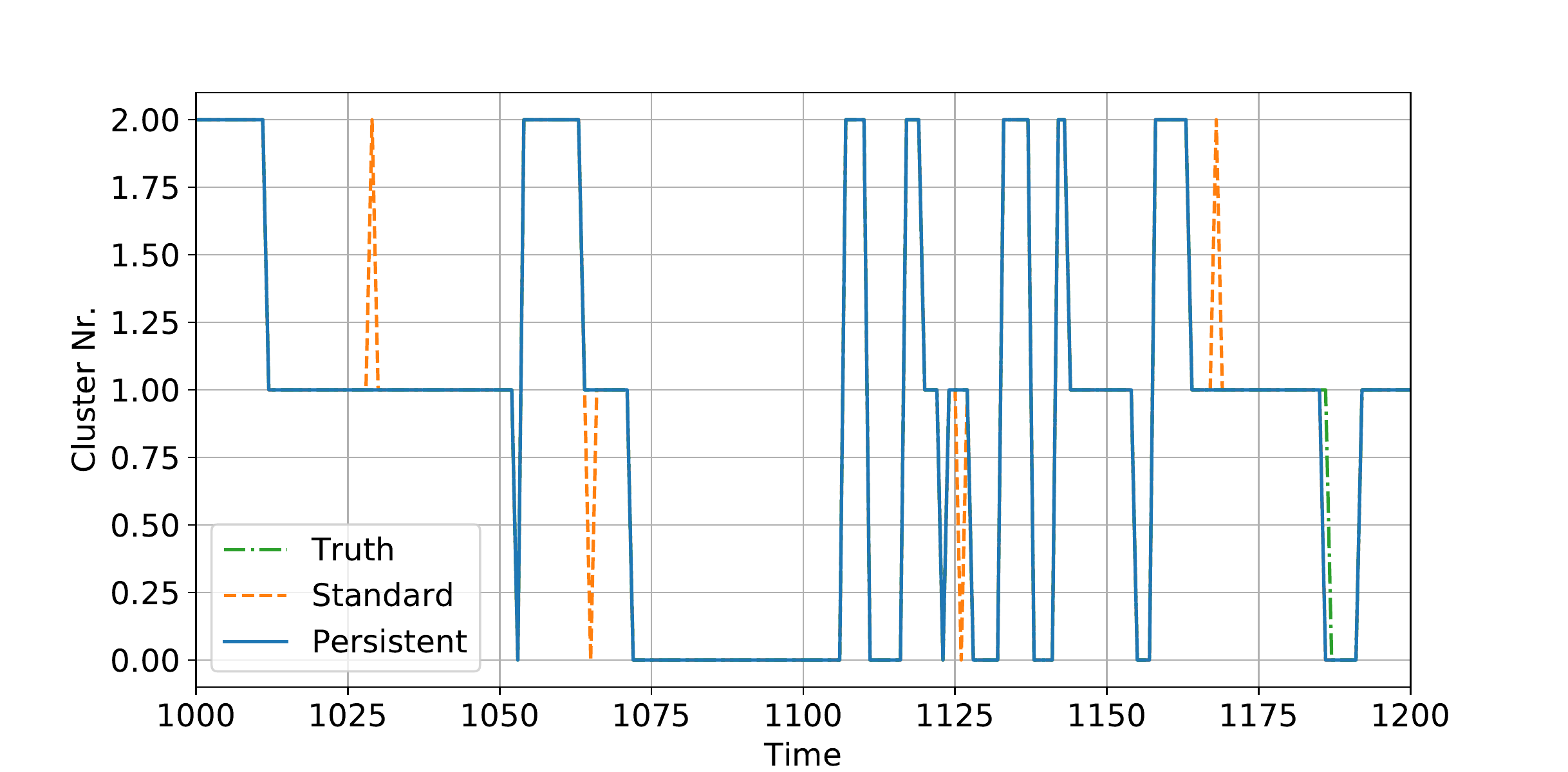}
		\caption{\label{fig:toyseq}A part of the transition sequence which indicates in which cluster the data belongs. The truth is shown in green (dash-dotted, visible on the very right), in orange (dashed) is the result for standard $k$-means and in blue (solid) the result when the persistence constraint is incorporated.  Note that the lines overlap most of the time.}
	\end{subfigure}
	\caption{\label{fig:toy}The toy model (a) showing the clusters in blue, green and purple and (b) the assignment of the data to the different clusters in time.}
\end{figure}

To make the effect of the incorporation of this constraint on the final clustering result more insightful a very simple toy model is presented (details given in the Supplementary Information). Consider a system of three 2D clusters which are normally distributed around their respective centres, each with a different variance. They transition into each other according to a persistent transition matrix, meaning there is a high probability of a cluster transitioning to itself. The clusters are shown in Figure \ref{fig:toymod}. When applying the standard (unconstrained) $k$-means algorithm the data that are assigned to the wrong cluster are located in-between the different clusters. In the assignment of the data to the clusters this leads to sudden jumps into one cluster and directly back to the original cluster, as can be seen in Figure \ref{fig:toyseq}, which leads to the identification of a too short persistence. When the persistence constraint is incorporated many of the wrongly assigned data points are now assigned to the correct cluster. Furthermore we see that the short transitions into and directly out of a cluster are removed. Thus the persistence found using the constrained algorithm is closer to that of the real system.

Because $k$-means clustering can only identify local minima we run the algorithm 500 times with different random initial conditions for the standard $k$-means algorithm. The initial condition at every location in space is drawn independently from a normal distribution around zero with the same standard deviation as the data. Note that this means there is no correlation in space, so as to not make any assumptions on the spatial patterns of the regimes. The tolerance used depends on $k$ and is $0.0001/k^2$ for the full field data and $0.001/k^2$ for the EOF data. The algorithm including the persistence constraint is run 100 times with different initial conditions; the reduced number of runs is due to increased computation time by the incorporation of linear programming, which is similar for the EOF and full field data. The final result is chosen to be the one with the smallest clustering functional $\mathbf{L}$. This method in general works reasonably well, but even for the Lorenz 63 system \cite[][]{Lorenz1963} the ``correct" clusters cannot be found for every realisation (simulation with different initial condition) when no persistence constraint is incorporated, as can be seen in the Supplementary Information. The data close to the cluster centres is always assigned correctly, but there is a substantial uncertainty in assigning the data further away from the cluster centres. Thus it is important to be careful when applying $k$-means clustering and not blindly trust the result, especially as the method assigns every data point to a cluster even if it actually is in-between different clusters. The incorporation of the persistence constraint improves this aspect of the $k$-means result.

\subsection{Information Criteria}
\label{ssec:meth_info}

For $k$-means clustering the number of clusters $k$ has to be set a priori and the question is how to determine the best value for $k$. The main methods used to this end in the identification of atmospheric circulation regimes are verification by synthetic datasets \cite[e.g.][]{Straus2007, Straus2017}, using a classifiability index \cite[e.g.][]{Michelangeli1995, Plaut2001}, and looking at the similarity of runs with different initial conditions \cite[e.g.][]{Jung2005}. An alternative method is to use an information criterion \cite[e.g.][]{OKane2013}, which is widely used in for example biological sciences \cite[e.g.][]{Volinsky2000, Posada2004, Arnold2010}. An information criterion is a tool from model selection which is used to identify the optimal model \cite[][]{Burnham2004}; it strikes a balance between how well the model fits the data and the number of parameters needed, to prevent over-fitting. The optimal balance is where the information criterion is minimal.  As the clusters are effectively a model representing the data, the concept can be applied here as well. This tool for identifying the optimal number of clusters has already been used in many applications and more theoretical studies \cite[e.g][]{Fraley1998, Chen1998, Cobos2014}. In addition to allowing to identify the optimal number of clusters $k$, an information criterion also allows for finding the best constraint value $C$ when persistence is incorporated in the clustering procedure. 

The two information criteria that are used most widely are the Akaike Information Criterion (AIC) and the Bayesian Information Criterion (BIC) \cite[][]{Burnham2004}. The AIC is based in information theory and is an approximation of how different two probability distributions (one for the data, one for the model) are \cite[][]{Akaike1973}. It is given by
\begin{equation}
\label{eq:AIC}
\text{AIC} = -2 \log(\mathcal{L}(\hat{\theta}| \text{data})) + 2 K, 
\end{equation}
where $\mathcal{L}(\hat{\theta}| \text{data})$ is the likelihood of the optimal model $\hat{\theta}$ given the data, which measures how well the model fits the data, and $K$ the number of parameters in the model. For cluster analysis the number of parameters is determined by the number of clusters, their dimension and the length of the data time series (expressions are given in the Supplementary Information). In contrast the BIC is based on the limiting behaviour of Bayes estimators, which minimizes the expectation value of the loss (e.g. error), and reads \cite[][]{Schwarz1978}:
\begin{equation}
\label{eq:BIC}
\text{BIC} = -2 \log(\mathcal{L}(\hat{\theta}| \text{data})) + K \log(n), 
\end{equation}
where $n$ is the sample size, here being the dimension of the data (number of gridpoints or number of EOFs) times the number of days. Just as with the AIC, the BIC strikes a balance between the (log-)likelihood, i.e. how well the clusters fit the data, and the number of parameters in the model. For both the AIC and BIC we refer to the second term as the penalty term since it penalizes the use of many parameters in finding the optimal model to prevent over-fitting.

To compute the values of both information criteria the log-likelihood is needed. Assuming the errors of the model are independent and normally distributed the log-likelihood term can be written as \cite[][]{Burnham2004}
\begin{equation}
\label{eq:loglik}
-2 \log(\mathcal{L}(\hat{\theta}| \text{data})) = n \log(\hat{\sigma}^2),
\end{equation}
where $\hat{\sigma}^2 = \sum \hat{\epsilon}_t^2 / n$ is the error variance for residuals $\hat{\epsilon}_t$, the latter being the difference between the cluster centres and the data for every grid point or EOF. This allows for a straightforward computation of both information criteria using the clustering functional.

The only difference between the AIC and the BIC is how they penalize the number of parameters in the model. The penalty term in the BIC takes into account the sample size, while the term in the AIC does not. This means the penalty term in the BIC is stronger with respect to the number of parameters, accounting for an assumed higher variability in a high-dimensional dataset, which increases the chances of over-fitting. Note that the number of parameters is different between the EOF and full field data since the number of parameters needed to identify a cluster centre is set by the dimension of the data. The dimension of EOF data is significantly smaller than that of full field data.
As a consequence the penalty term of the BIC is stronger relative to the log-likelihood term for EOF data compared to full field data. This is illustrated in Table S2 of the Supplementary Information using 20 EOFS. For this reason the BIC is expected to not perform well for EOF data, by which we mean that it will identify a very low number of clusters $k$ to be optimal. On the other hand, the penalty term in the AIC likely is too weak to yield a realistic optimal $k$ for the full field data. A high number of clusters is expected to be found optimal, well beyond what is physically reasonable and suitable. When using either of these criteria one always has to judge whether the result is sensible for the purpose of the study.

\section{Results}
\label{sec:res}

First we compare the results of the regimes found using the EOF data and the full field data. In this we focus on the optimal number of regimes using the information criteria discussed. Second we discuss the difference between enforcing persistence in the clustering method and the use of time-filtered data on the occurrence and persistence of the regimes.

\subsection{Number of Regimes}
\label{ssec:res_numb}

The standard number of wintertime regimes identified over the Euro-Atlantic sector in literature is four \cite[e.g][]{Vautard1990, Cassou2008, Dawson2015}. Few studies question this number \cite[e.g.][]{Fereday2008}. This optimal number of four clusters has always been found in the context of EOF data. The most-used argument for four regimes being optimal is based on how consistent, or similar, the results of the $k$-means algorithm are for different (random) initial conditions. A way in which this is commonly assessed is the classifiability index introduced by \citet[][]{Michelangeli1995}, which uses pattern correlation to determine how similar two sets of regimes are. The value of the index is computed for both the data and a synthetic dataset with the same statistics, to see whether there exists a $k$ for which the result is significantly different from a noise model. Another method to assess the similarity is to look at the assignment of the data to the different regimes \cite[][]{Fereday2008}. This similarity measure considers the regimes for all data (instead of only their aggregate) and thus possibly provides more information than the pattern correlation.

\begin{figure}
	\centering
	\begin{subfigure}{.95\textwidth}
		\centering
		\includegraphics[width=1.\textwidth]{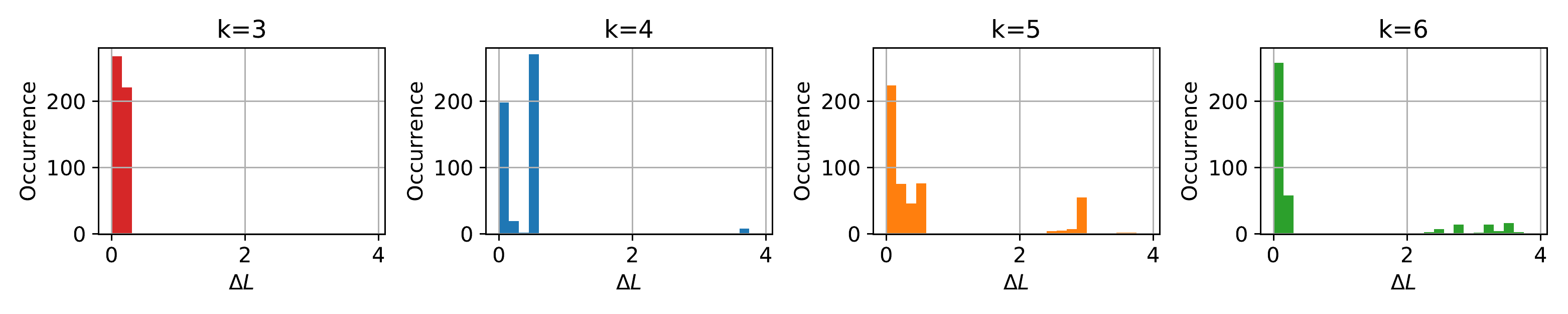}
		\vspace{-20pt}
		\caption{\label{fig:L_PCA_k36}The clustering functional $\mathbf{L}$.}
	\end{subfigure}
	\begin{subfigure}{.95\textwidth}
		\centering
		\includegraphics[width=1.\textwidth]{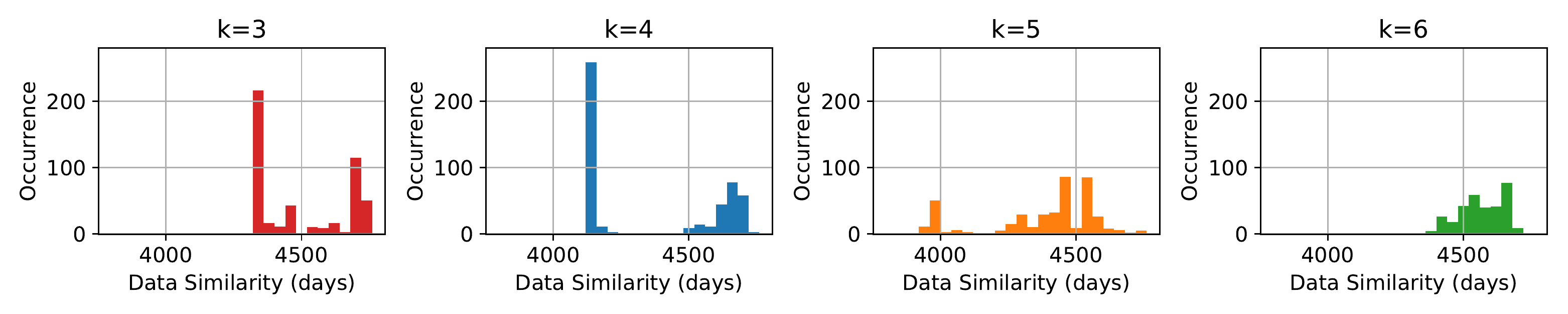}
		\vspace{-20pt}
		\caption{\label{fig:Data_PCA_k36}The data similarity.}
	\end{subfigure}
	\caption{\label{fig:sim_PCA_k36}Histograms for the difference of the clustering functional with its minimum value $\Delta\mathbf{L} = \mathbf{L}_{\text{run}} - \mathbf{L}_{\text{min}}$ and the data similarity with respect to the optimal (minimal $\mathbf{L}$) result using the first 20 EOFs for $k=3,...,6$.}
\end{figure}

Here we briefly examine the data similarity, as well as the spread in the clustering functional $\mathbf{L}$, for the EOF regimes with the main aim being to verify the reliability of the found regimes. Histograms for $\Delta\mathbf{L} = \mathbf{L}_{\text{run}} - \mathbf{L}_{\text{min}}$, being the difference of $\mathbf{L}$ for each run with the minimal value over all runs of the clustering algorithm with different initial conditions, and the number of days assigned to the same regime (as the best result) are shown in Figure \ref{fig:sim_PCA_k36}. The first aspect to note is that for some $k$ not only a global minimum is found, but also a local one (e.g. for $k=4$ in Figure \ref{fig:L_PCA_k36}). In addition we see that the found regimes can be quite different in their assignment of data to certain regimes, indicating the found regimes are significantly different. Instead of only combining all accurate results (near the global minimum of $\mathbf{L}$) by computing the average data similarity (similar to the classifiability index \cite{Michelangeli1995}) we also look at its variance. The values for $k=3,...,6$, given on the left side of Table \ref{tab:measure}, show that the variance is lowest for $k=4$, which goes together with a high average. This is consistent with the results found in literature. Interestingly the mean for $k=6$ is slightly higher than found for $k=5$, while the opposite would be expected as more clusters allow for more variability.

\begin{table}
	\caption{\label{tab:measure}The mean ($\mu$) and variance per cluster ($\sigma^2/k$) of the data similarity for data with $\mathbf{DL} = \mathbf{L}_{i} - \mathbf{L}_{i+1}$, where the $\mathbf{L}_i$ are sorted from small to large, below a set threshold for both the EOF and full field results. For the full field results also the values for the odd and even years are given. The number of runs that are below the threshold is shown as well.}
	\centering
	\small{
	\begin{tabular}{c | ccc || ccc | ccc | ccc}
		& \multicolumn{3}{c||}{20 EOFs: $\mathbf{DL}<0.01$} & \multicolumn{9}{c}{Full field: $\mathbf{DL}<0.0005$} \\
		& \multicolumn{3}{c||}{All years} & \multicolumn{3}{c|}{All years} & \multicolumn{3}{c|}{Odd years} & \multicolumn{3}{c}{Even years} \\
		$k$ & $\mu$ & $\sigma^2/k$ & \#data & $\mu$ & $\sigma^2/k$ & \#data & $\mu$ & $\sigma^2/k$ & \#data & $\mu$ & $\sigma^2/k$ & \#data \\
		\hline
		3 & 4643 & 3649 & 254 & 4552 & 2109 & 485 & 2342 & 875 & 409 & 2246 & 140 & 156 \\
		4 & 4658 & 330 & 197 & 4607 & 1440 & 201 & 2359 & 452 & 248 & 2255 & 37 & 423 \\
		5 & 4509 & 978 & 265 & 4660 & 149 & 204 & 2187 & 3564 & 274 & 2243 & 132 & 137 \\
		6 & 4571 & 1103 & 315 & 4581 & 790 & 316 & 2296 & 322 & 60 & 1911 & 10478 & 210 \\
		7 & - & - & - & 3686 & 37440 & 417 & 1998 & 12145 & 165 & 2113 & 1470 & 170
	\end{tabular}
}
\end{table}

\begin{figure}
	\centering
	\begin{subfigure}{.95\textwidth}
		\centering
		\includegraphics[width=1.\textwidth]{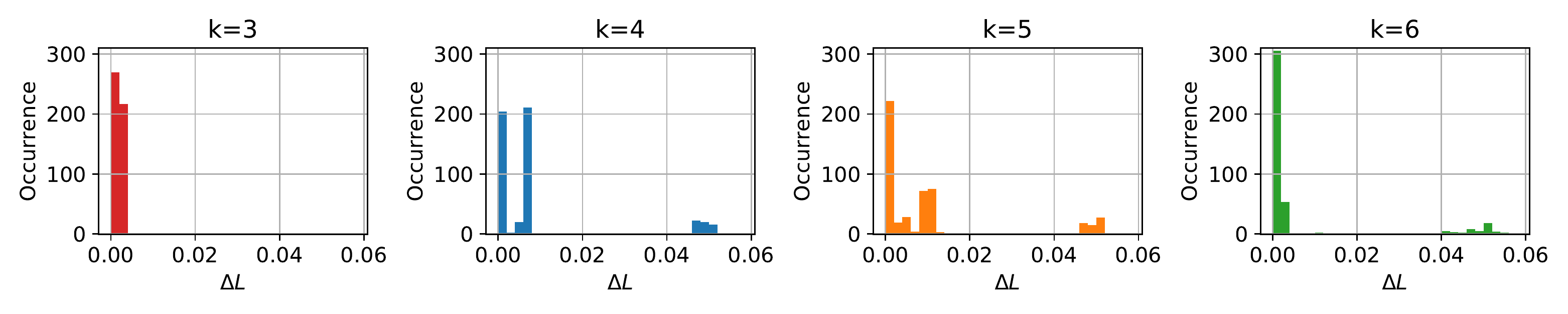}
		\vspace{-20pt}
		\caption{\label{fig:L_k47}The clustering functional $\mathbf{L}$.}
	\end{subfigure}
	\begin{subfigure}{.95\textwidth}
		\centering
		\includegraphics[width=1.\textwidth]{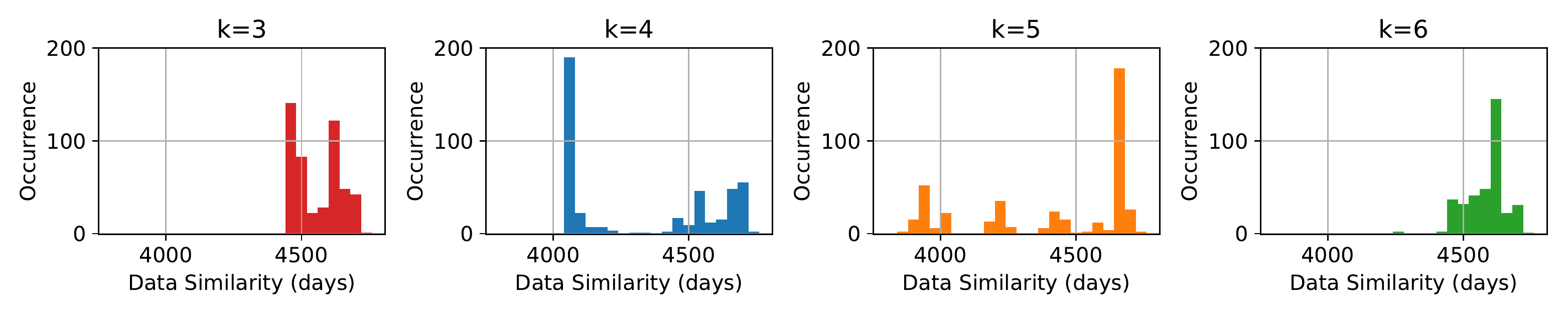}
		\vspace{-20pt}
		\caption{\label{fig:Data_k47}The data similarity.}
	\end{subfigure}
	\caption{\label{fig:sim_k47}Histograms for the difference of the clustering functional with its minimum value $\Delta\mathbf{L} = \mathbf{L}_{\text{run}} - \mathbf{L}_{\text{min}}$ and the data similarity with respect to the optimal (minimal $\mathbf{L}$) result using the full field data for $k=3,...,6$.}
\end{figure}

Next we turn to the distribution of $\Delta\mathbf{L}$ and the data similarity for the full field data, as shown in Figure \ref{fig:sim_k47}. The first thing to note is that the distributions of $\Delta\mathbf{L}$ look similar to those for the EOF data, indicating that using the high dimensional full field data does not reduce the chance of finding the optimal regimes. Some differences with the EOF result do occur for the data similarity, most notably the increased similarity for $k=5$. Looking at the results for the global minimum we find that both $k=5$ and $k=6$ show a smaller variance than $k=4$, in contrast to the EOF results. Especially for $k=5$ the difference is substantial. These differences indicate that by performing an EOF analysis some information is lost, resulting in a stronger consistency of $k$-means for $k=5$ and $k=6$ when using the full field data.
	
In addition we look at the results of the clustering algorithm for a subset of the data, which is a standard approach to test the robustness of clustering methods to e.g. identify cats on photos \cite[][]{Jain2010}. Here the results for the datasets of odd and even years are studied, as stationarity of the dataset cannot be assumed. The differences between the results for odd and even years are found to be large, indicating either $k=4$ or $k=6$ having the smallest variance. Also for $k=5$ differences between the two sets of years are large. These ambiguous results raise the question whether half the dataset is of sufficient length to draw reliable conclusions about the clustering results. This also means that non-stationarity of the regimes due to e.g. climate change is difficult to study accurately using clustering methods applied to reanalysis datasets.

\begin{figure}
	\centering
	\begin{subfigure}{.49\textwidth}
		\centering
		\includegraphics[width=1.\textwidth]{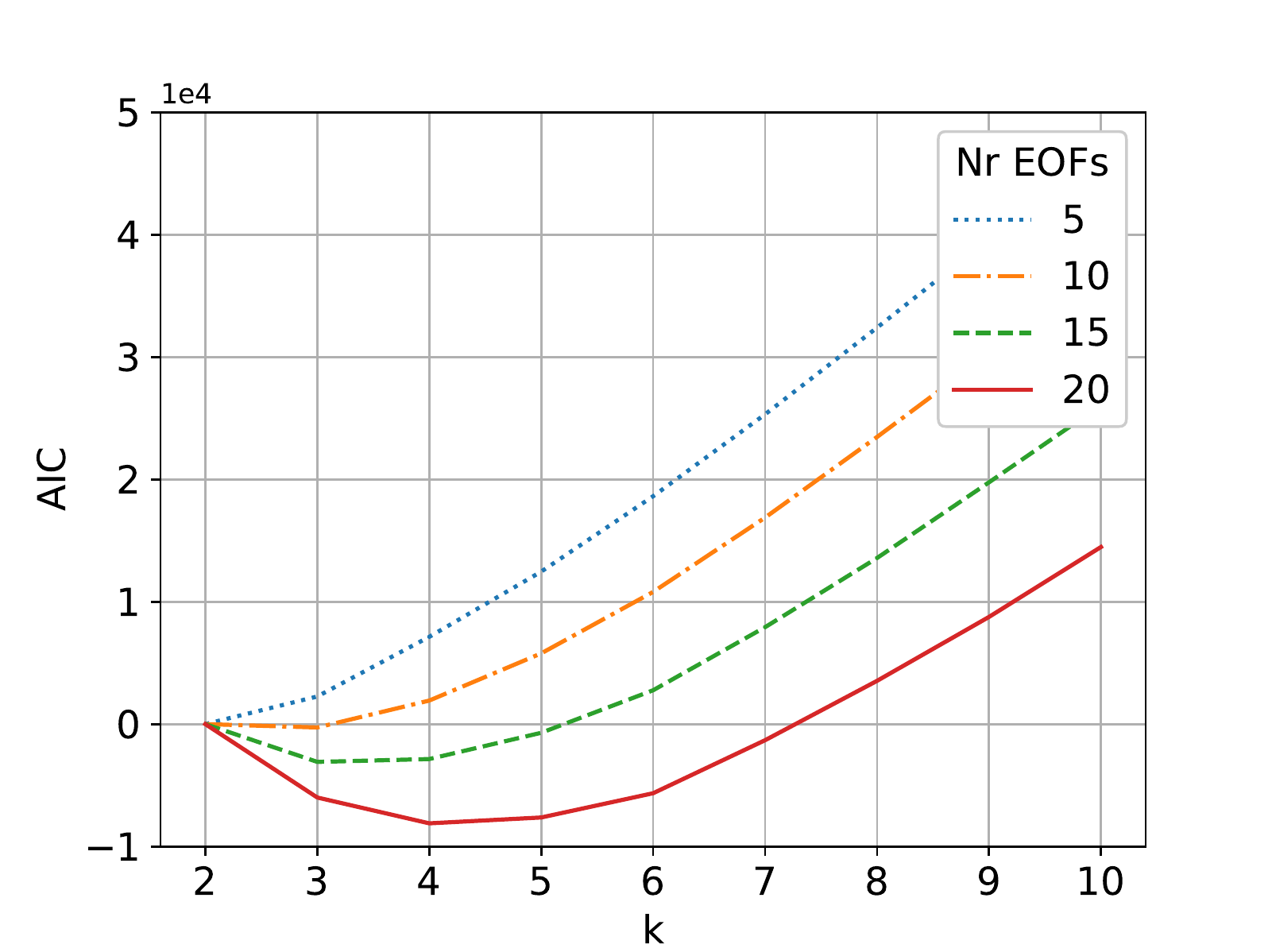}
		\caption{\label{fig:IC_EOF}The AIC for different numbers of EOFs.}
	\end{subfigure}
	\begin{subfigure}{.49\textwidth}
		\centering
		\includegraphics[width=1.\textwidth]{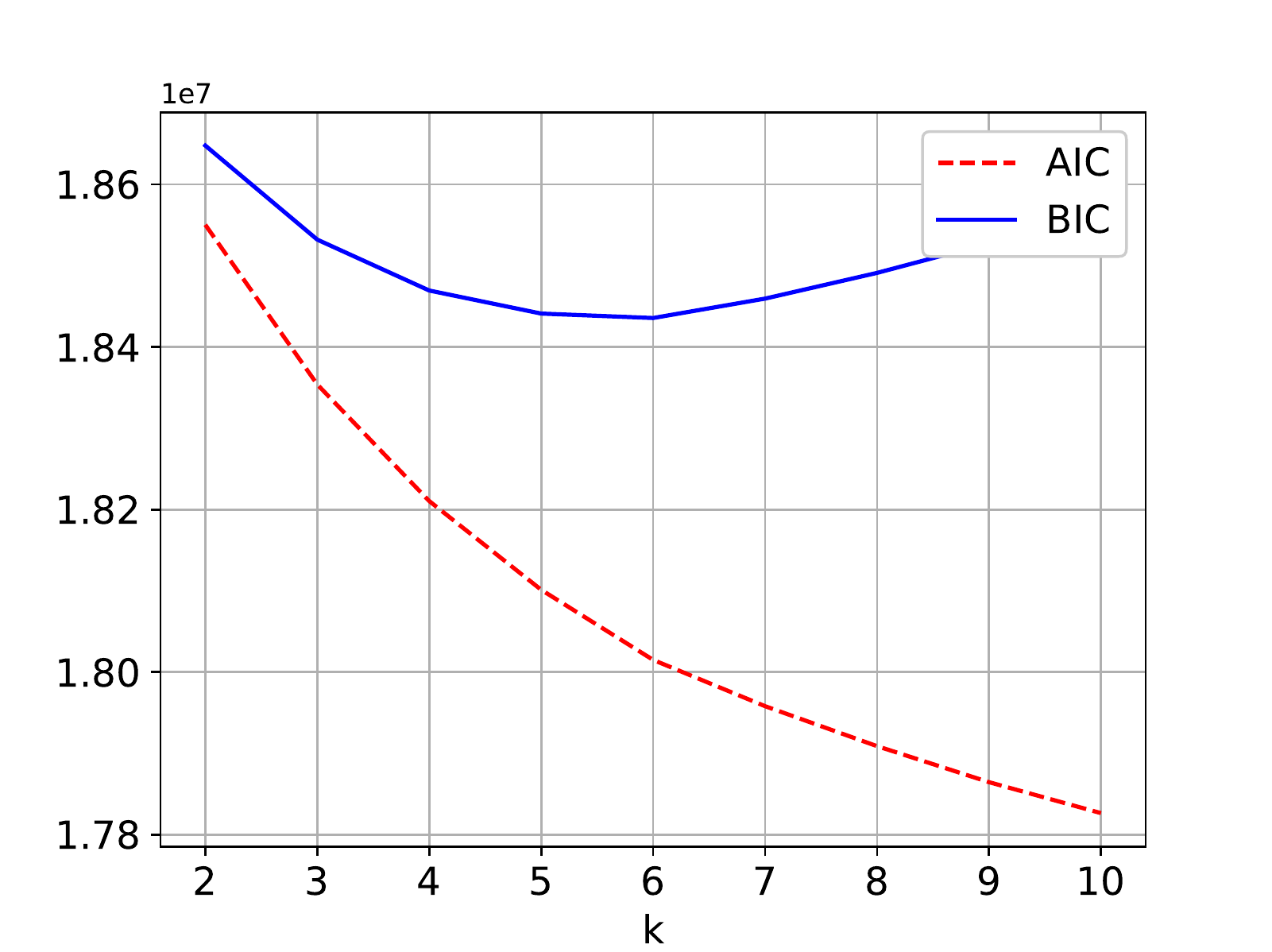}
		\caption{\label{fig:IC_field}The AIC and BIC for the full field data.}
	\end{subfigure}
	\caption{\label{fig:IC}Information criteria for both the full field and EOF datasets for the range $k = 2,...,10$. Both the AIC and BIC, as given in Section \ref{ssec:meth_info}, are shown for the full field data. For the EOF data only the AIC is shown.}
\end{figure}

We refrain from drawing definite conclusions about the optimal number of clusters $k$ from the above discussion on consistency, as there is some debate about its suitability for this purpose \cite[][]{Philipp2007}. Instead, we use the AIC and BIC to identify the optimal number of regimes. The AIC is used when considering the EOF results, as it is expected to give better results in that case as discussed in Section \ref{ssec:meth_info}. In Figure \ref{fig:IC_EOF} the AIC is shown for using 5, 10, 15 and 20 EOFs to identify the circulation regimes. A minimum at $k=4$ is found when 20 EOFs are used, although the AIC is also small for $k=3$ and $k=5$. For lower numbers of EOFs the optimal number is found to be lower, while a higher number of EOFs leads to a higher optimum for $k$. This is to be expected because the use of a limited number of EOFs means that some variability of the original data is neglected. This loss of variability is larger when less EOFs are used and as a consequence fewer clusters are needed to account for the variability of the EOF data. The BIC has its minimum at $k=2$ for every number of EOFs considered, indicating the penalty term for the number of parameters is indeed too strong for the EOF data (Section \ref{ssec:meth_info}). Based on the AIC we conclude that $k=4$ is indeed the optimal number when using 20 EOFs, which corresponds with results from literature. However, for other numbers of EOFs the optimal $k$ according to the AIC is different, meaning this conclusion is not unambiguous.

The optimal number of regimes identified by an information criterion when using the full field dataset is not the same. The BIC is more suitable for the full field data than the AIC due to the dependence of the penalty term on the sample size, making it stronger than the penalty term of the AIC (Section \ref{ssec:meth_info}). In Figure \ref{fig:IC_field} both the AIC and BIC are shown for the full field data. The BIC points towards an optimum of $k=6$. The AIC does not show a minimum in the range considered as the penalty term is not strong enough for the high dimensional full field data. Therefore, we base our decision on the optimal number of regimes on the BIC and find $k=6$ to be optimal.

\begin{figure}
	\centering
	\includegraphics[width=.9\textwidth]{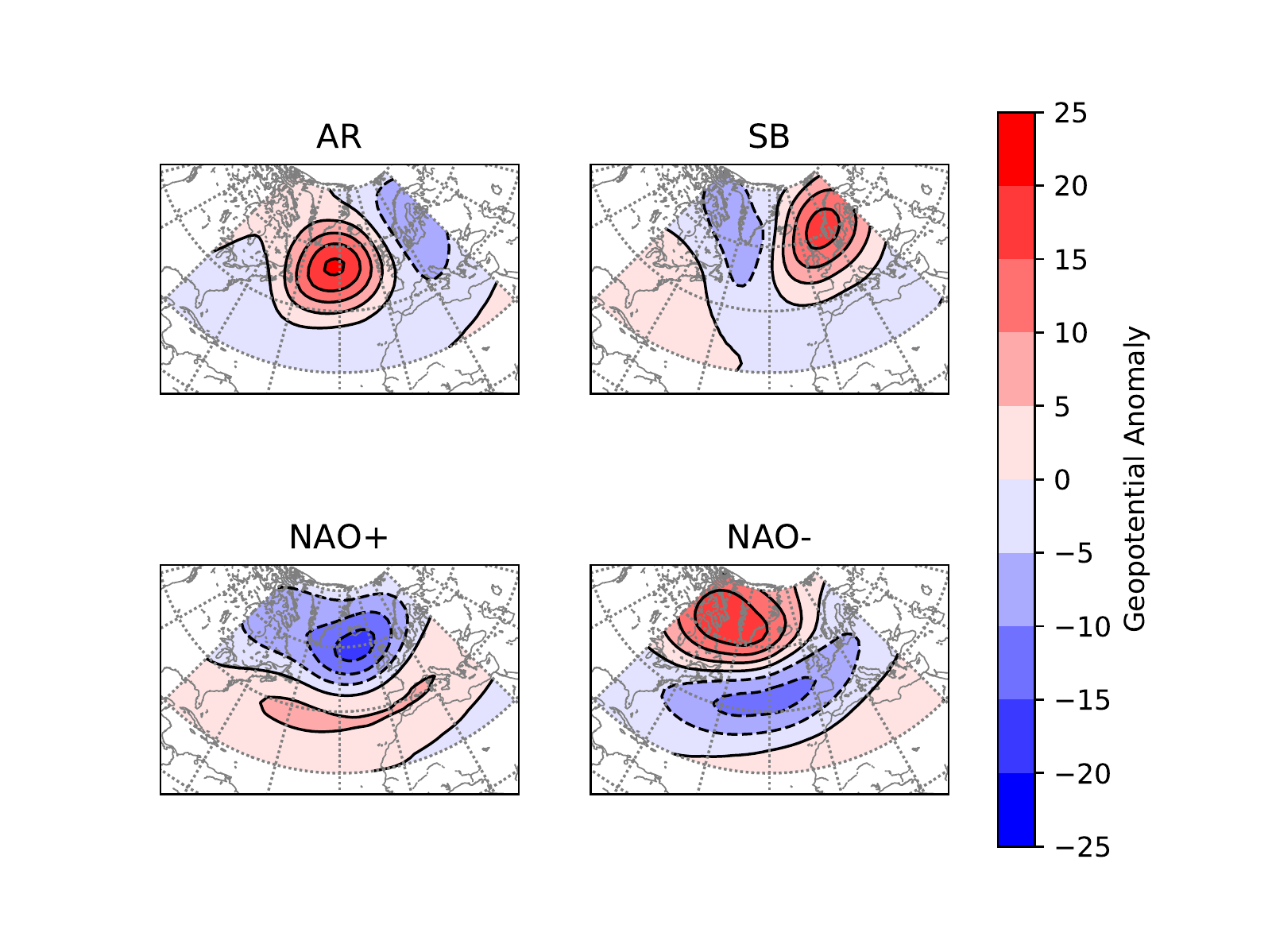}
	\vspace{-18pt}
	\caption{\label{fig:regimes_k4}The clustering result of the standard $k$-means algorithm applied to the full field data for $k=4$.}
\end{figure}

The regimes that are obtained by using 10 or more EOFs and the full field data do not differ substantially for the same $k$. Similarly the occurrence rate and transition probabilities of the regimes do not differ substantially. In Figure \ref{fig:regimes_k4} the four regimes known from literature \cite[e.g.][]{Hannachi2012, Straus2007} are shown as obtained by applying $k$-means clustering on the full field data for $k=4$. They are the Atlantic Ridge (AR), Scandinavian Blocking (SB) and the two phases of the North Atlantic Oscillation (NAO). The transition probability of a regime to itself (or daily re-occurrence rate) and overall occurrence rate of these regimes can be found in Table \ref{tab:occpersval}. The positive phase of the NAO is the most frequently occurring regime, followed by SB. The high occurrence of the NAO+ regime may reflect the fact that it is the only regime associated with a northern low pressure area. Both phases of the NAO are found to be most persistent, i.e. transition to themselves with the highest probability, while the AR exhibits the least persistence. We note that the occurrence rates obtained are similar to those found in literature despite not using a seasonally varying background state.

\begin{figure}
	\centering
	\includegraphics[width=.9\textwidth]{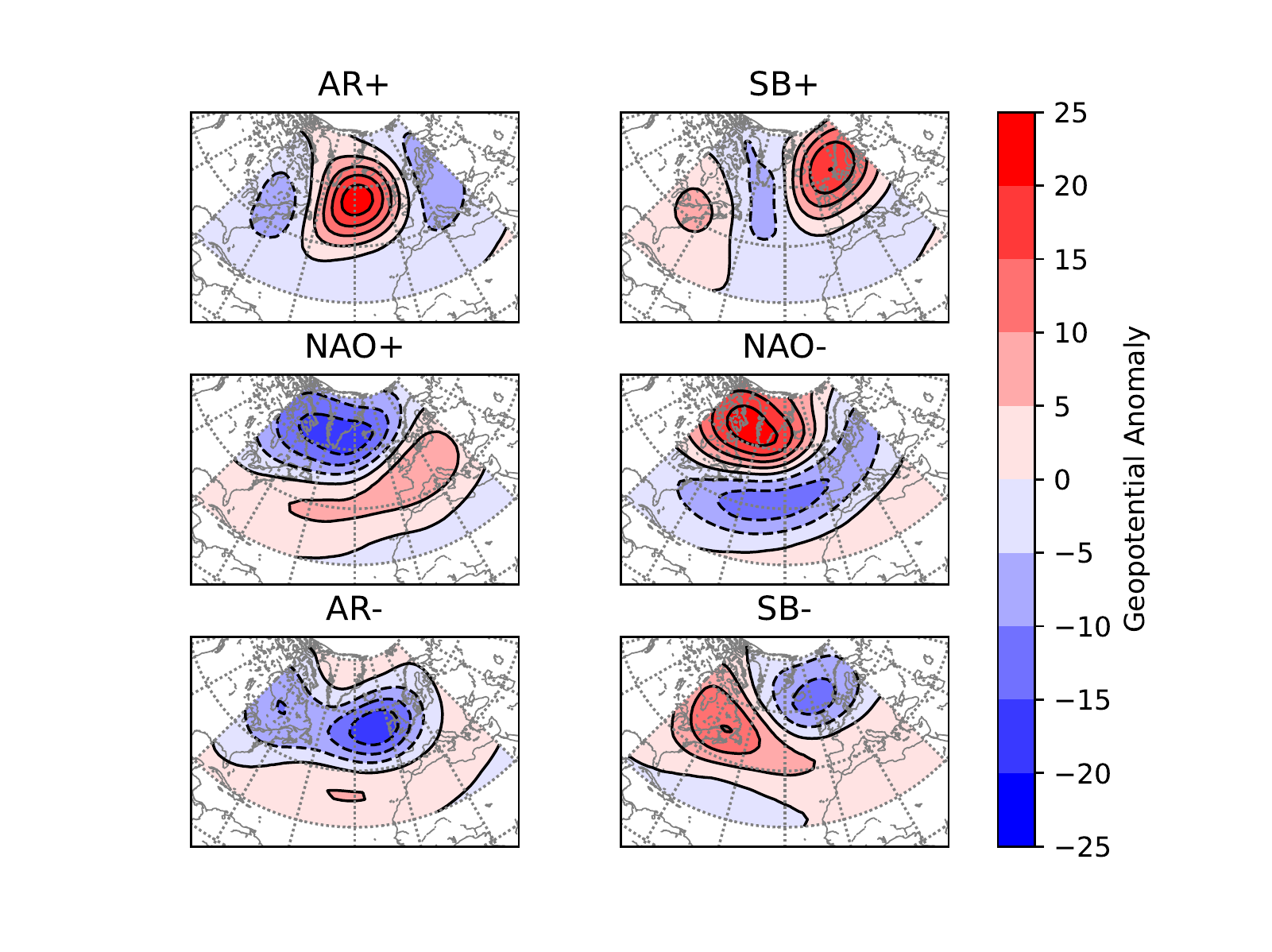}
	\vspace{-18pt}
	\caption{\label{fig:regimes_k6}The clustering result of the standard $k$-means algorithm applied to the full field data for $k=6$.}
\end{figure}

In Figure \ref{fig:regimes_k6} the regimes found using $k$-means clustering on the full field data for $k=6$ are shown. The first four regimes are in essence the same as those found for $k=4$ in Figure \ref{fig:regimes_k4}. Small differences occur in the location of the maximum high or low pressure area for the AR and NAO+. The two additional regimes found have a low pressure area either in the central Atlantic or over Scandinavia. The first thus can be identified as the opposite phase of the AR with a pattern correlation of -0.57 and we refer to it as AR-, while the original regime is denoted by AR+. Similarly we refer to the second additional regime as SB-, as it represents the opposite phase of the SB regime (from now on denoted by SB+), where the pattern correlation -0.49 is slightly lower. Note that the pattern correlation of the two phases of the NAO is higher for $k=6$, with the value being -0.93, versus -0.59 when using $k=4$. The use of six clusters thus introduces a pleasing symmetry in the found regimes, with an equal number of regimes having a high and low pressure area in the north of the domain.

The occurrence rate and self-transition probability of the six regimes show different behaviour than found for $k=4$, as can be seen in Table \ref{tab:occpersval}. Instead of the NAO+ the SB+ is found to be the most frequently occurring regime. The NAO+ is the second ranked regime in occurrence, albeit with a small but significant difference relative to SB+. The other four regimes show similar occurrence rates. When looking at the self-transition probabilities, the NAO- remains the most persistent, with exactly the same probability. The NAO+ however does lose some of its persistence, reducing its self-transition probability to a rate similar to that of the SB+. The AR- is found to be the second most persistent regime and the AR+ remains the least persistent. In Table \ref{tab:occpersval} the e-folding time and average regime duration computed using the self-transition probability are also given for reference. Furthermore, we note that the occurrence rate and self-transition probabilities change throughout the winter months. Figures showing this seasonal behaviour are given in the Supplementary Information. This variation could reflect the effect of the seasonal cycle in the data, as well as intrinsic variability in the regime behaviour.

\begin{table}
	\caption{\label{tab:occpersval} The values of the occurrence rate, self-transition probability, e-folding time and average regime duration for the unconstrained result for both $k=4$ and $k=6$.}
	\centering
	\small{
	\begin{tabular}{l | cccc | cccccc}
		& \multicolumn{4}{c|}{$k=4$} & \multicolumn{6}{c}{$k=6$} \\
		& AR & SB & NAO+ & NAO- & AR+ & SB+ & NAO+ & NAO- & AR- & SB- \\
		\hline
		Occurrence (\%) & 21.3 & 26.8 & 31.5 & 20.4 & 15.6 & 19.6 & 16.9 & 15.5 & 16.3 & 16.1 \\
		Self-Transition Probability & 0.756 & 0.792 & 0.850 & 0.849 & 0.712 & 0.748 & 0.751 & 0.847 & 0.787 & 0.730 \\
		e-Folding Time (days) & 3.6 & 4.3 & 6.2 & 6.1 & 2.9 & 3.4 & 3.5 & 6.0 & 4.2 & 3.2 \\
		Average Duration (days) & 4.1 & 4.8 & 6.7 & 6.6 & 3.5 & 4.0 & 4.0 & 6.5 & 4.7 & 3.7
	\end{tabular}
}
\end{table}

\subsection{Persistence}
\label{ssec:res_pers}

In Section \ref{ssec:meth_clus} two methods to enforce persistence of the atmospheric circulation regimes have been discussed. The first method is the standard approach of applying a time-filter, and the second method is to include a persistence constraint in the clustering algorithm itself. Note that the average regime duration of averaged over the unconstrained regimes following Table \ref{tab:occpersval} is 5.5 days for $k=4$ and 4.4 days for $k=6$, which means the persistence constraint will only affect the result when it is below either $C\approx1800$ or $C\approx2200$ for $k=4$ or $k=6$ respectively (Table \ref{tab:Cday}). The regimes found using these two methods do not differ substantially from those found and discussed in the previous section. When a time-filter is applied to the data the regimes are found to be slightly weaker, meaning the maximum and minimum geopotential height anomaly are smaller, but they do not show a visible difference in the configuration of high and low pressure areas. For the results using a persistence constraint differences in the regimes only emerge for very strong (unrealistic) constraints in the form of slight shifts in the location of the centres of high and low pressure. For weak (realistic) constraints the regimes found are the same as for the unconstrained method and no weakening is found. By a `realistic' constraint we mean one that does not force data points into regimes which are a large distance away, but only switches those data points that are in-between different regimes, as can be seen in the toy example in Section \ref{ssec:meth_clus}. In practice these are constraints corresponding to an average regime duration below circa 9 days (the corresponding $C$ can be found in Table \ref{tab:Cday}).

\begin{figure}[h]
	\centering
	\begin{subfigure}{.49\textwidth}
		\centering
		\includegraphics[width=1.\textwidth]{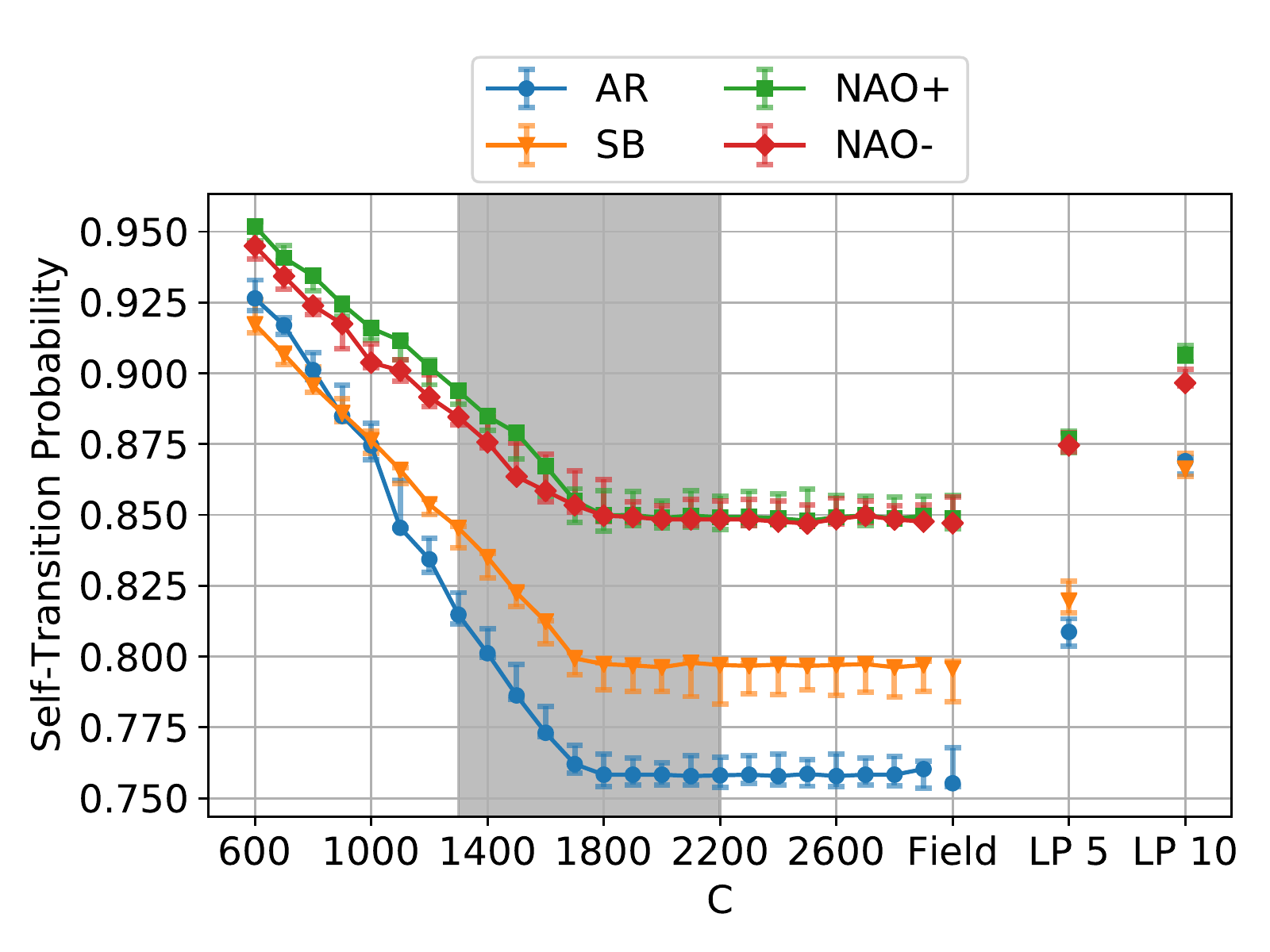}
		\vspace{-18pt}
		\caption{\label{fig:pers_C_k4}The self-transition probabilities for $k=4$.}
	\end{subfigure}
	\begin{subfigure}{.49\textwidth}
		\centering
		\includegraphics[width=1.\textwidth]{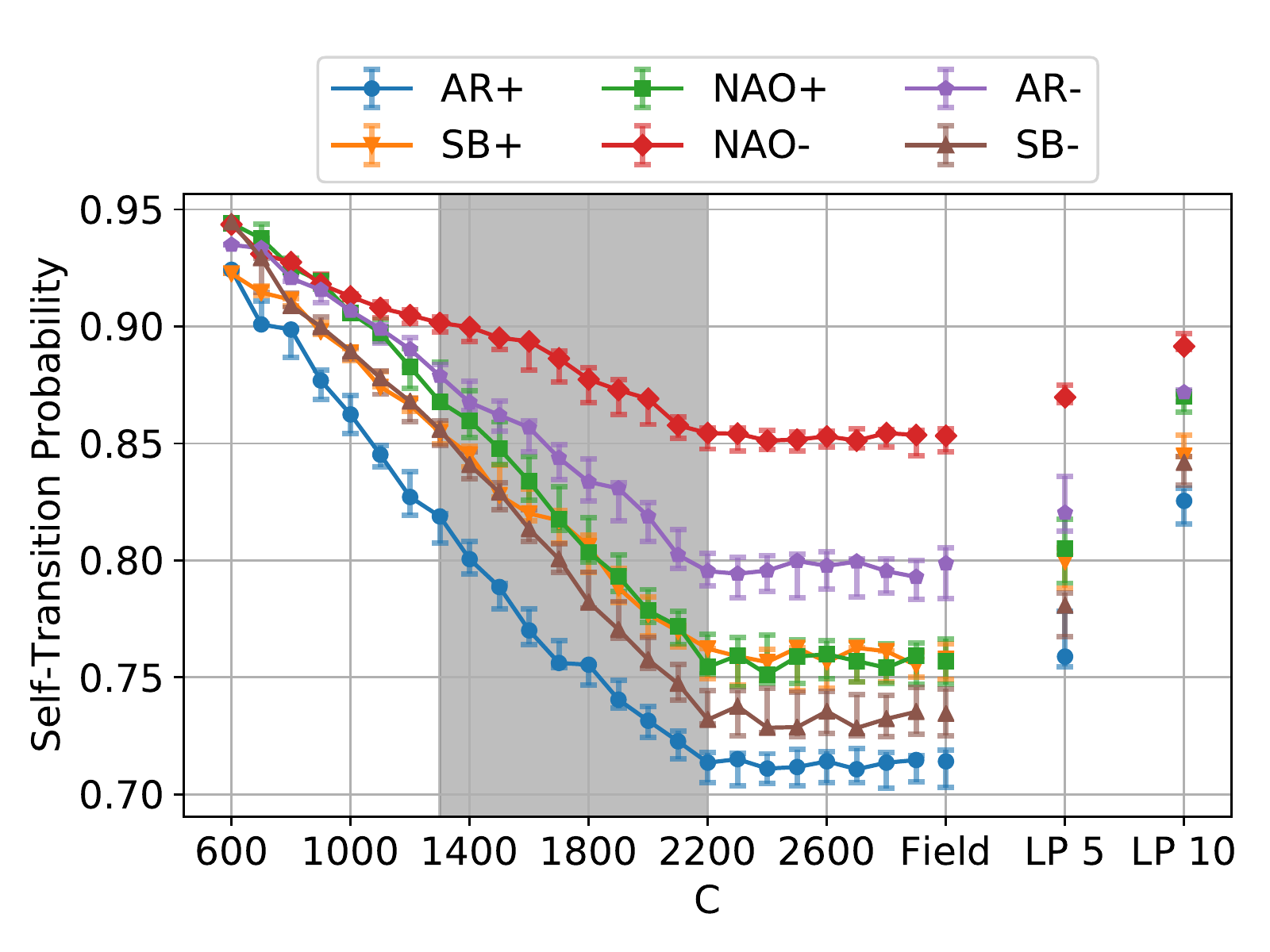}
		\vspace{-18pt}
		\caption{\label{fig:pers_C_k6}The self-transition probabilities for $k=6$.}
	\end{subfigure}
	\begin{subfigure}{.49\textwidth}
		\centering
		\includegraphics[width=1.\textwidth]{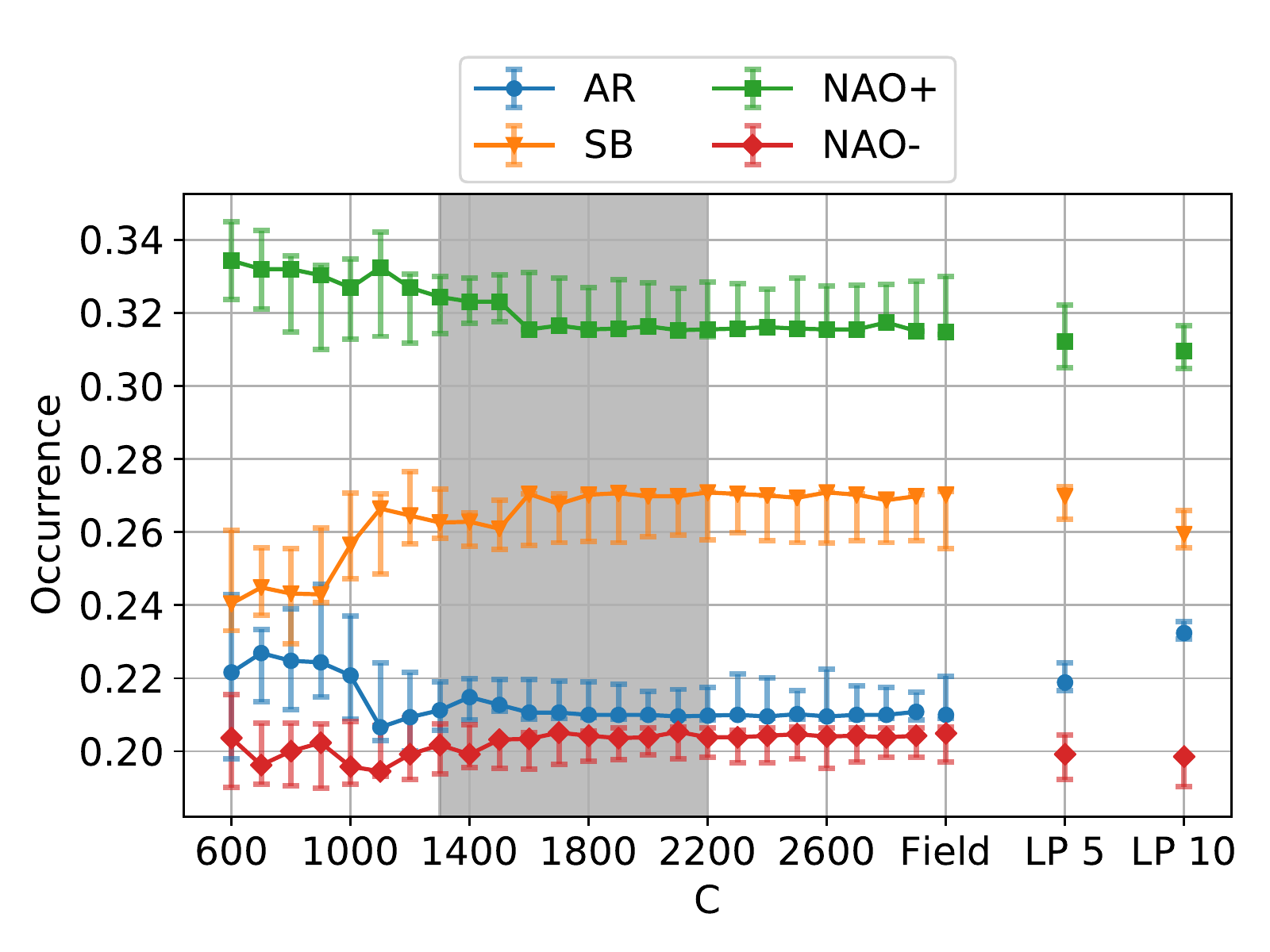}
		\vspace{-18pt}
		\caption{\label{fig:occ_C_k4}The occurrence rates for $k=4$.}
	\end{subfigure}
	\begin{subfigure}{.49\textwidth}
		\centering
		\includegraphics[width=1.\textwidth]{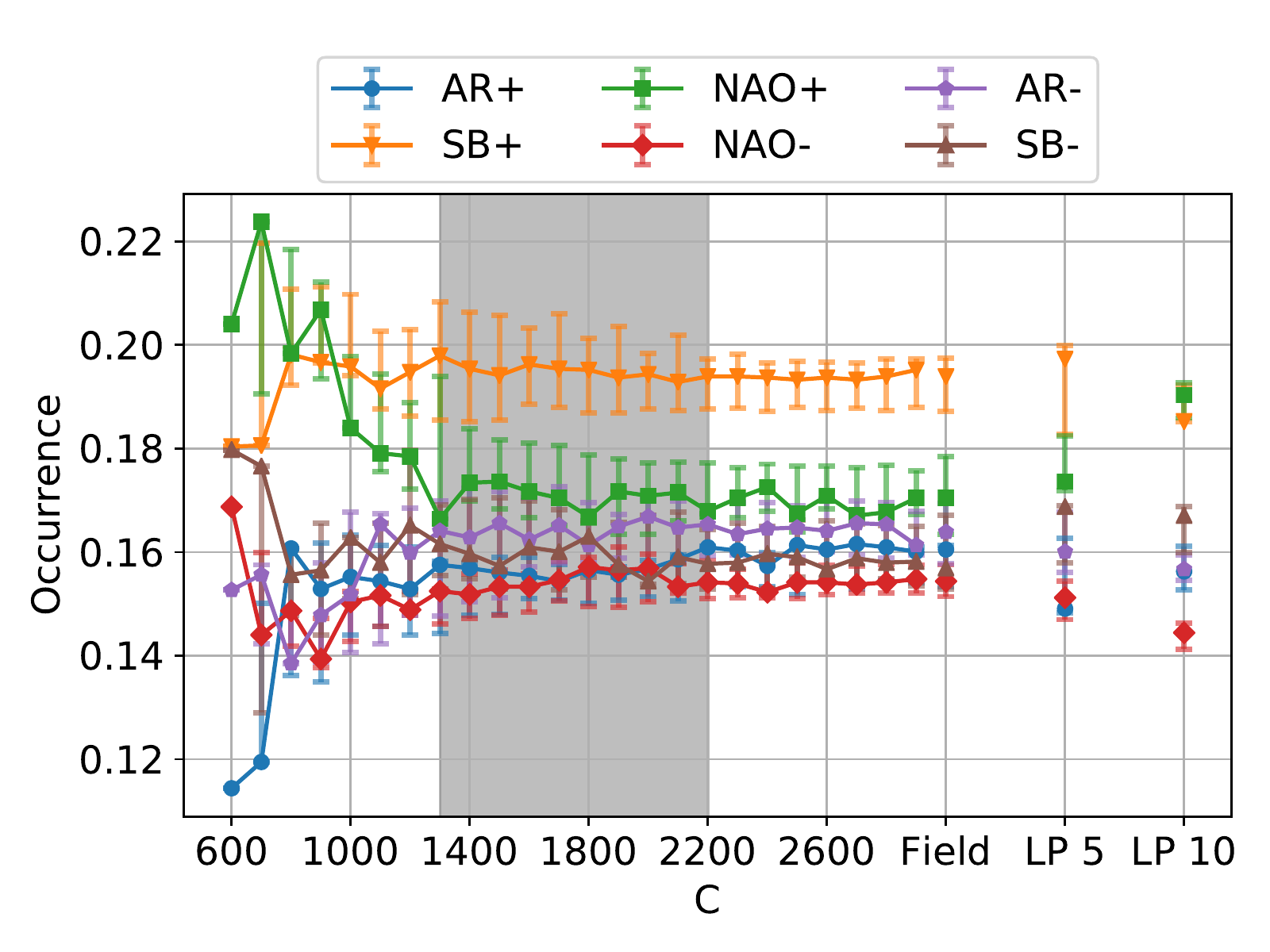}
		\vspace{-18pt}
		\caption{\label{fig:occ_C_k6}The occurrence rates for $k=6$.}
	\end{subfigure}
	\caption{\label{fig:occpers_C}The occurrence and self-transition probabilities of the different regimes for $k=4$ and $k=6$ for the clustering results including the persistence constraint depending on the value of $C$ (corresponding average regime durations can be found in Table \ref{tab:Cday}). To the right the values for the unconstrained algorithm (field) and the 5- and 10-day low-pass filter are shown. The error bars indicate the maximum and minimum value of occurrence/self-transition probabilities for clustering results with a slightly smaller $\mathbf{L}$ (bounds for the difference are \{0.00968, 0.00936, ..., 0.00232, 0.002, 0.002, 0.002\} decreasing with increasing $C$, which are chosen sufficiently small to give similar regimes according to the data correspondence. The gray bands indicate the region in which the persistence constraint is considered to act as a good filter.}
\end{figure}

The effects of the time-filtering and persistence constraint method on the self-transition probability are shown in Figures \ref{fig:pers_C_k4} ($k=4$) and \ref{fig:pers_C_k6} ($k=6$). On the left of each panel the results for the constrained algorithm are shown for various $C$ and as expected the self-transition probability increases with decreasing $C$. The smaller the value of $C$, the less switches between regimes are allowed. The increase of the self-transition probability with decreasing $C$ is approximately linear for all regimes, and starts at the `raw' self-transition probability of the regimes. Consistently with the values computed using the average regime duration in Table \ref{tab:occpersval} we find that the constraint starts to affect the self-transition probabilities around either $C=2200$ for $k=6$ (4.3 days) or $C=1800$ for $k=4$ (5.3 days). Note that when the constraint starts to affect the results the regime dynamics can no longer be described as a first-order Markov process and care must be taken interpreting the average regime duration and corresponding self-transition probabilities.

Comparing the results for time-filtered data with those of the constrained method in Figure \ref{fig:occpers_C} we see that using a 5-day low-pass filter corresponds to a constraint of roughly 2000 for $k=6$ and 1400 for $k=4$. This difference is mainly due to the stronger effect of the constraint for a larger number of clusters. For the 10-day filter the corresponding values of $C$ are approximately 1400 and 1100 for $k=6$ and $k=4$ respectively. Note that the self-transition probability of certain regimes differs slightly between the two methods. For example the AR+ regime is found to increase its self-transition probability relatively stronger for the time-filtered data.

The occurrence rates of the different regimes are shown in Figures \ref{fig:occ_C_k4} and \ref{fig:occ_C_k6} for $k=4$ and $k=6$ respectively. We start by looking at the results of the constrained algorithm. The occurrence rate remains the same as for the unconstrained data, even for constraint values significantly stronger than the `raw' persistence of the data. Only for very low $C$ (strong constraints) do the occurrence rates start to differ. This indicates that the method indeed causes a switching of the `in-between-cluster' points to the cluster of their neighbours instead of the cluster they are slightly closer to, as expected from the results of the toy model (Section \ref{ssec:meth_info}). We regard the constraint as being `weak' so long as the occurrence rates are not affected, and helping to identify true physical persistence. For these values of $C$ we consider that the persistence constraint acts as a good filter, indicated by the gray bands in Figure \ref{fig:occpers_C}. In contrast, the results for the time-filtered data show significant differences in the occurrence rates of the regimes. Especially for the 10-day filter the differences for e.g. the AR+ regime ($k=4$) or the NAO+ regime ($k=6$) are substantial. As this is the standard filter used in literature \cite[e.g.][]{Straus2017} it raises the question of how reliable the occurrence rates are and whether they are not solely a feature of the method used. In contrast, the inclusion of the constraint within the clustering procedure itself does not lead to such a bias and therefore provides a more robust way of finding persistent regimes, i.e. of isolating the signal from the noise.

\begin{figure}
	\centering
	\includegraphics[width=.7\textwidth]{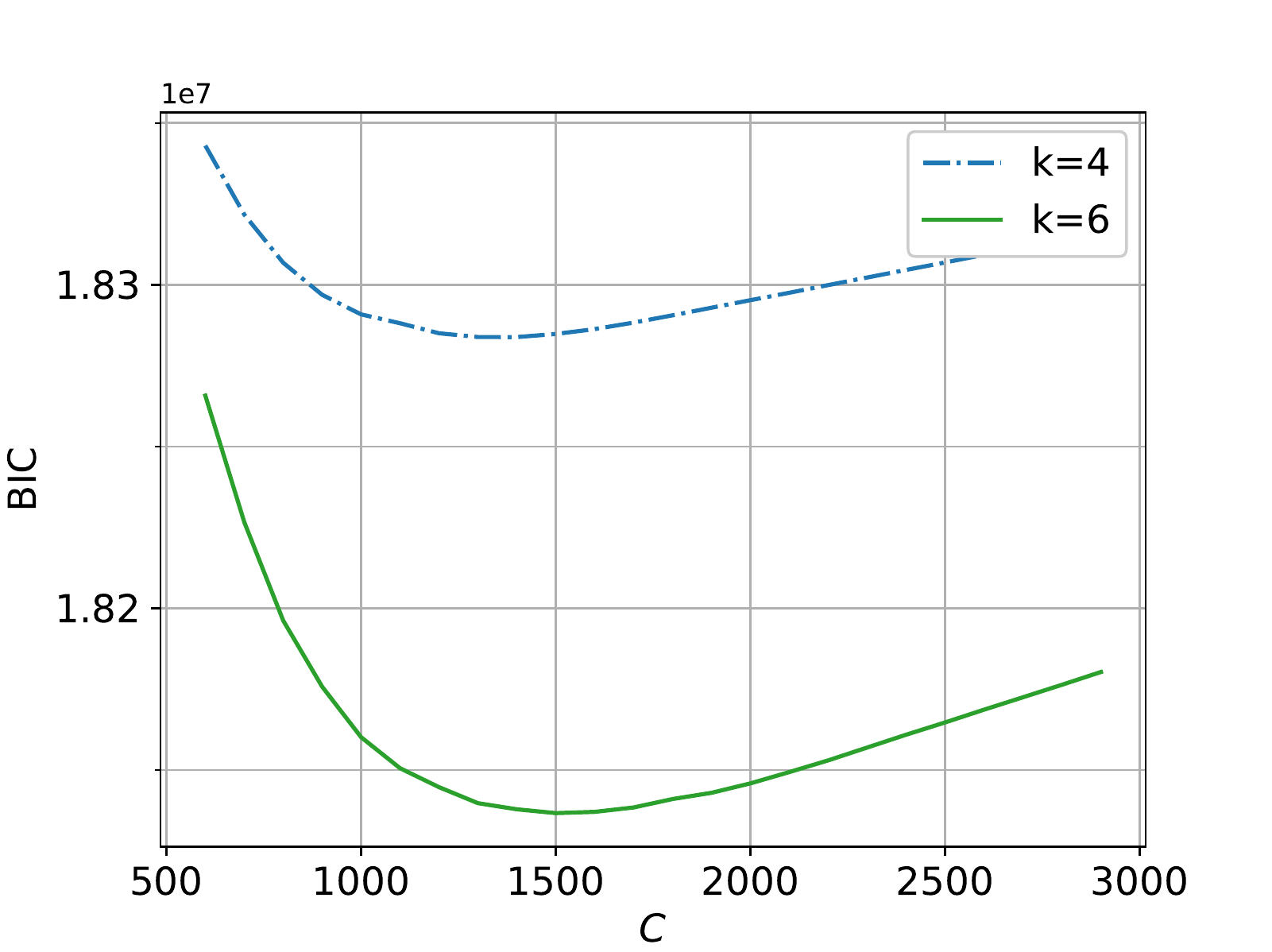}
	\caption{\label{fig:IC_pers}The BIC for the clustering with persistence constraint for $k=4,6$ and $C$ in the range $\{600,..., 2900\} $.}
\end{figure}

When using the constrained algorithm one of the choices that needs to be made is which constraint value $C$ is best to use. Here we base this choice on the BIC, as we did for finding the optimal number of clusters $k$. In Figure \ref{fig:IC_pers} the BIC is shown for both $k=4$ and $k=6$. For $k=4$ the minimum is found for $C=1400$ and for $k=6$ it is found for $C=1500$. These constraint values correspond to an average regime duration of 6.8 and 6.3 days respectively. The lower end of the region in which the BIC is close to its minimum coincides with the point beyond which smaller values of $C$ start to affect the occurrence rates, giving a lower bound for the region where the persistence constraint is considered to act as a good filter as indicated by the gray bands in Figure \ref{fig:occpers_C}.  This increases the confidence of the optimal value of $C$ being around these values. Interestingly, the optimal average regime duration for $k=4$ and $k=6$ differs by less than 10\% ($\Delta C \approx 1500 - 1400 = 100$), whereas without the persistence constraint the average duration differs by 20\% ($\Delta C \approx 2200 - 1800 = 400$). This confirms that the persistence constraint is helping identify a physical signal that is less dependent on the number of clusters chosen. The range where the BIC is very close to its minimum is between 6.3 ($C=1500$) and 7.9 ($C=1200$) days for $k=4$. For $k=6$ this range is from 5.9 ($C=1600$) to 6.8 ($C=1400$) days. Twice this timescale, which is the minimum for recurrence of a regime (i.e. for regimes $A$ and $B$ we have $A - B - A$), thus corresponds roughly to a period of 12 to 14 days. This is somewhat longer than the timescale of synoptic weather systems \cite[][]{Blackmon1977, Boljka2018a}. As this is an average there are a substantial number of longer lasting regimes showing persistence well beyond the synoptic timescale.

\section{Conclusion and Discussion}
\label{sec:disc}

In this study we have shown, using an information criterion and further arguments based on the consistency of the clustering result, that the traditional number of four clusters is not optimal for representing wintertime Euro-Atlantic weather regimes when full field data is used. The traditional approach of applying clustering to the first few EOFs involves a loss of information, which affects the number of regimes that is best to represent the data. The optimal number of regimes for the full field data was identified using the Bayesian Information Criterion (BIC), which strikes a balance between how well the regimes fit the data and the number of parameters needed to describe them.

We find that for the full field data, six regimes is the optimal choice. The two additional regimes are the opposite phases of the Atlantic Ridge and Scandinavian Blocking, introducing a pleasing symmetry in the found clusters. Furthermore, the dominant occurrence of the NAO+ when there are only four clusters, which likely is due to it being the only regime with a low pressure area in the north, is reduced by the addition of two regimes that also have this feature. Therefore, six regimes allow for more variability in their representation of the circulation and prevent all data with a more zonal flow from projecting onto the NAO+.

Furthermore we looked into ways to enforce persistence of the regimes. A common approach in literature is to apply a low-pass filter to remove high frequency oscillations and focus on the persistent behaviour \cite[e.g.][]{Bao2015, Straus2017}. This alters the data to which the clustering algorithm is applied, just as the use of EOFs does. We have shown that this leads to a significant change in the occurrence rate of the circulation regimes. A new method, which incorporates a persistence constraint in the algorithm itself, does not change the data while still enforcing persistent regimes. The results for this approach do not exhibit the change in occurrence rate found for the time-filtered data, as long as the constraint is not too strong, while still having an increased self-transition probability. Therefore this method leads to a more robust and unbiased result compared to the time-filtering approach.

A choice that needs to be made in this adapted clustering method is the value of the constraint $C$. Using the BIC the optimal value of $C$ is found to lie around an average regime duration of six to seven days. Interestingly, this matches the point beyond which smaller values of $C$ start to affect the occurrence rates. Thus it can be viewed as a more accurate estimate of the physical persistence of the regimes than that provided by the raw data without the persistence constraint. Double this value, which is the minimum for recurrence of a regime, thus is slightly longer than that of synoptic weather systems \cite[][]{Blackmon1977, Boljka2018a}. This shows that the atmospheric circulation indeed exhibits persistence beyond the synoptic timescale, suggesting the presence of predictable low-frequency modes.

Both results indicate that care must be taken when applying filtering methods (EOFs, low-pass filter) to the data before a clustering algorithm is applied. Clustering itself provides a means of dimension reduction, by projecting onto components representing recurrent patterns in the data. Since this is a method of filtering the data it seems ill-advised to apply this to already filtered EOF data, as it is not clear what the effect of this double filtering is on the result. A similar argument holds for applying a time-filter to the data before clustering. Information is lost in this procedure, introducing a bias in the resulting circulation regimes and their occurrence rates.

\section{Acknowledgements}
\label{acknowledgements}

SF was supported by the Centre for Doctoral Training in Mathematics of Planet Earth, UK EPSRC funded (grant EP/L016613/1). TGS and JdW were supported by ERC Advanced Grant ``ACRCC” (grant 339390). The research of JdW has been partially funded by Deutsche Forschungsgemeinschaft (DFG) - SFB1294/1 - 318763901 and by the Simons CRM Scholar-in-Residence Program. ERA‐Interim data are publicly available at the ECMWF website (http://apps.ecmwf.int/datasets/data/interim-full-daily/levtype=pl/). We thank David Straus and an anonymous reviewer for their thoughtful comments.

\section{Supporting Information}
\label{suppinfo}

Supporting Information is available online. It contains a discussion of the effects of seasonality on the regimes, technical details of the time-filter, the toy model, the information criteria and relation between the e-folding time and the average regime duration, as well as additional figures of distributions of $\Delta L$ and the data similarity, and clustering results for the Lorenz 63 system.

\selectlanguage{english}
\FloatBarrier

\clearpage

{\centering \Huge{Supporting Information}\\[15mm] \par}

\appendix
\beginsupplement

\section{Seasonality in the Identification of Atmospheric Circulation Regimes}
\label{sec:season}

\subsection{Seasonality in the Background Circulation}
\label{sec:background}

\begin{figure}[h]
	\centering
	\begin{subfigure}{.48\textwidth}
		\centering
		\includegraphics[width=1.\textwidth]{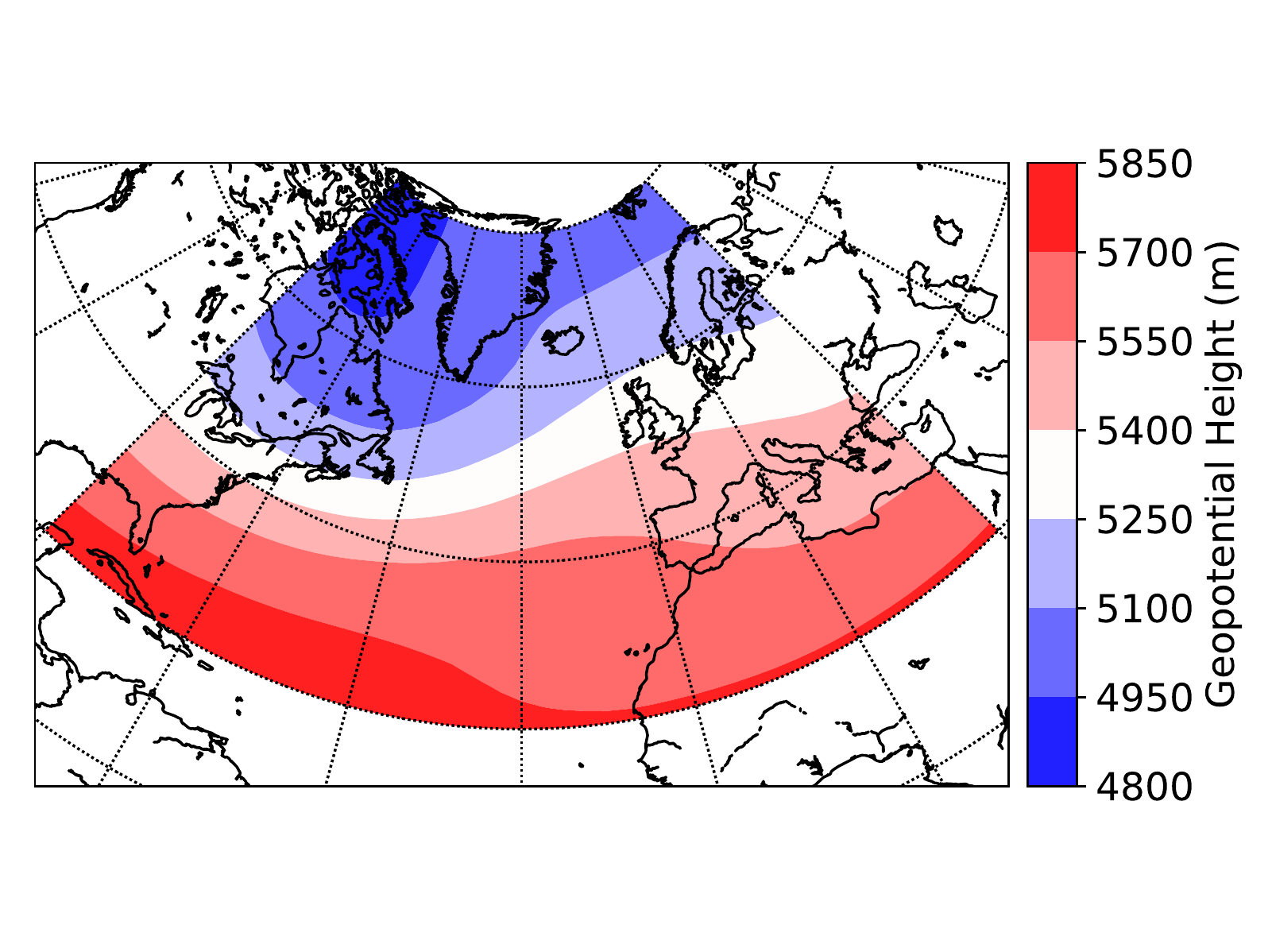}
		\vspace{-25pt}
		\caption{\label{fig:janav}The average geopotential height for January over all years considered.}
	\end{subfigure}
	\hspace{10pt}
	\begin{subfigure}{.48\textwidth}
		\centering
		\includegraphics[width=1.\textwidth]{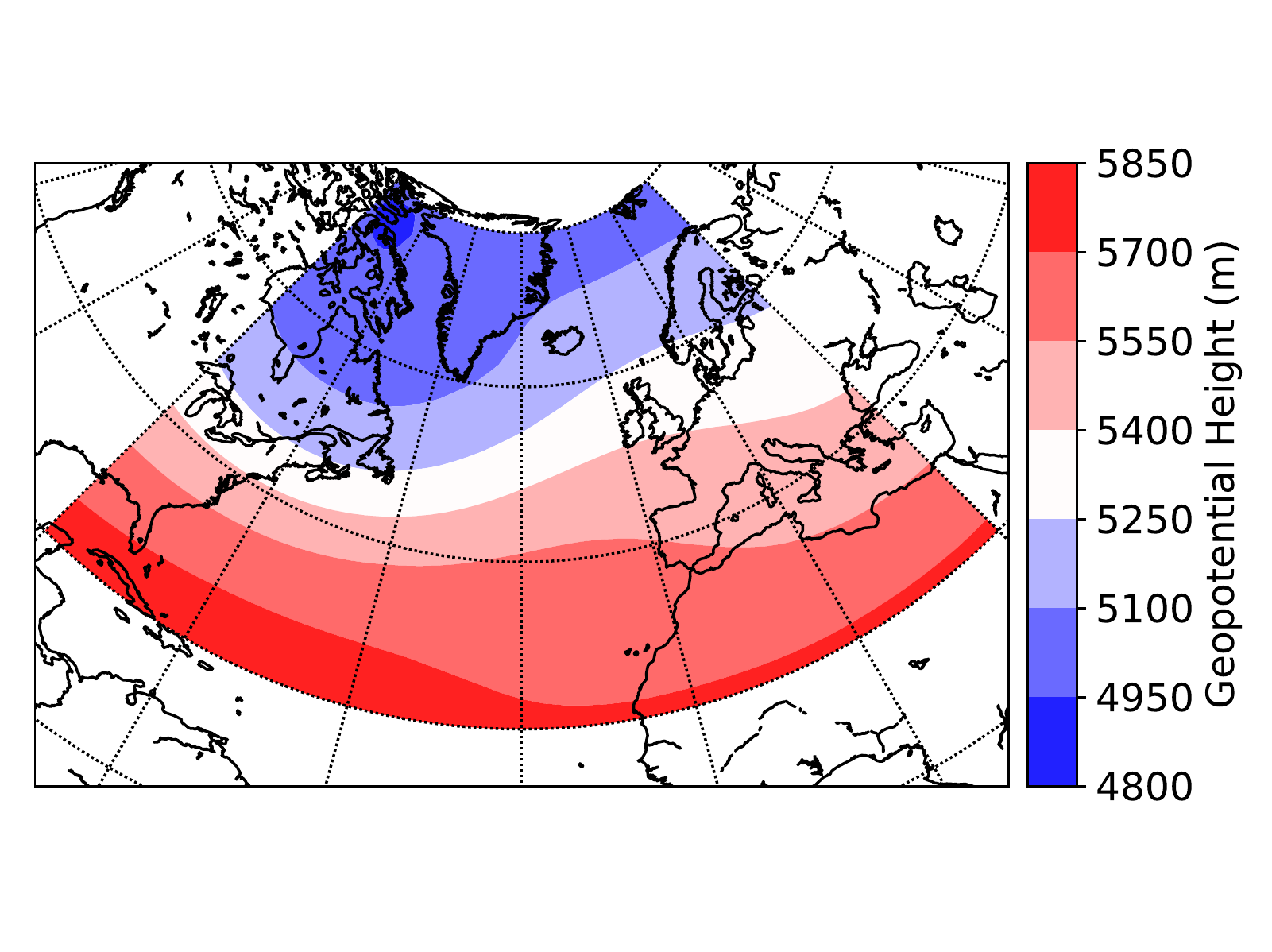}
		\vspace{-25pt}
		\caption{\label{fig:background} The average geopotential height for December through March over all years considered.}
	\end{subfigure}
	\begin{subfigure}{1.\textwidth}
		\centering
		\includegraphics[width=.9\textwidth]{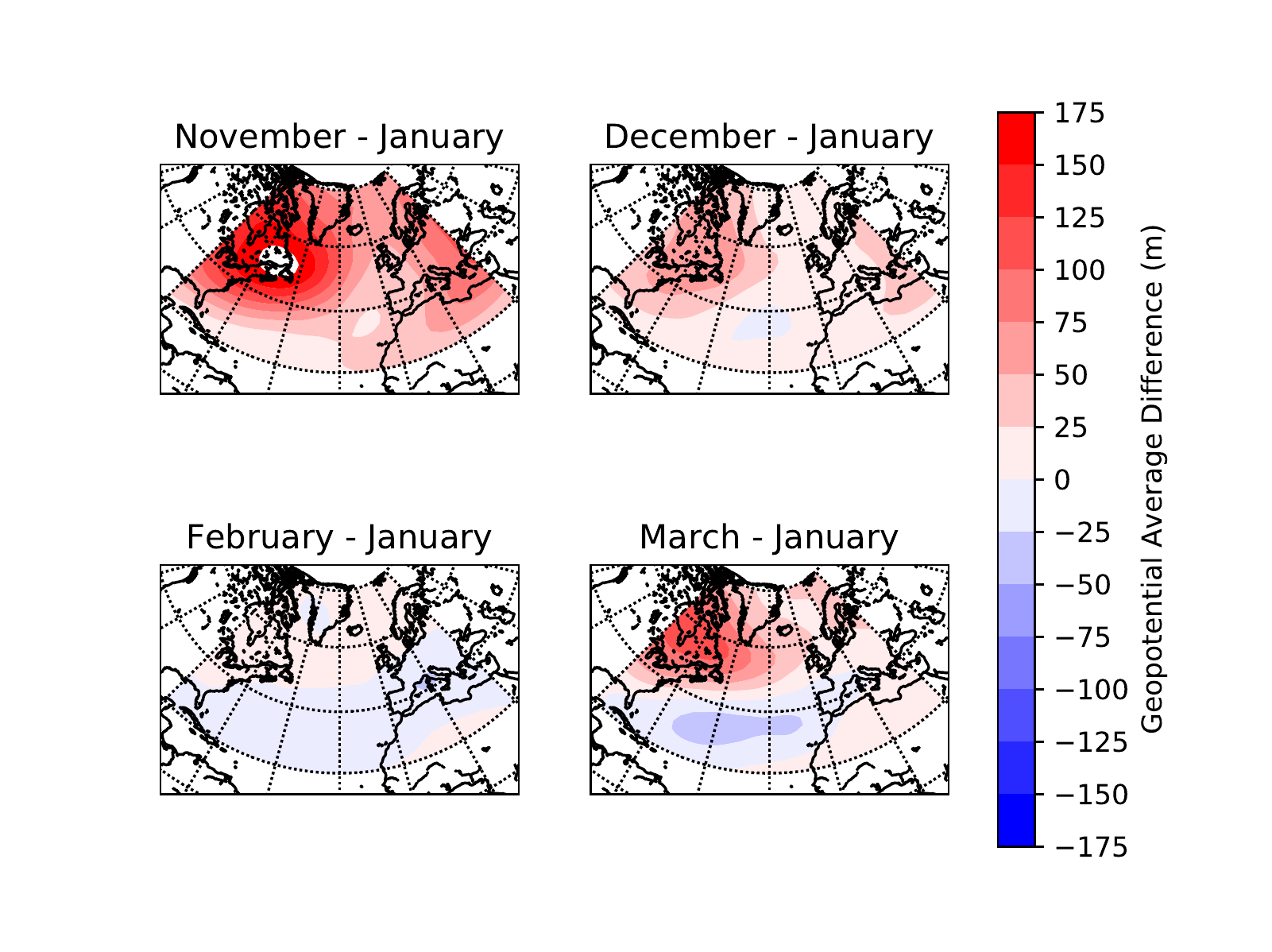}
		\vspace{-20pt}
		\caption{\label{fig:jandiff}The difference in average geopotential height between January and each of the other months.}
	\end{subfigure}
	\caption{\label{fig:seasonback}}
\end{figure}

We apply the clustering method to deviations from a fixed background state. The choice of which months to take into account for this background state can have a considerable impact on the resulting regimes and thus needs to be subject to careful consideration. Therefore we look in detail at the average geopotential height for the months November through March. The centre of winter is January so we take this month as a reference (Figure \ref{fig:janav}). We then compare the average state of the other months with that of January to assess when seasonality becomes important.

The differences between the average geopotential height for all months with January are shown in Figure \ref{fig:jandiff}. Clearly the difference with November is the largest, while that with February is very small. To make this comparison more quantitative we compute the distance between the average geopotential height of the different months with that of January in the cosine-weighted $L^2$-norm. The values are shown in the first row of Table \ref{tab:bgdiff}. February is by far the closest to January, so taking the average of these two months as a background state is not expected to result in a strong effect of seasonality. In the other rows of Table \ref{tab:bgdiff} the difference between the average over periods of several months and the months separately is given. Clearly November is the outlier with the largest distance to the average even if it is included in the months considered. Therefore it is expected that including November in the period over which the average is computed will lead to significantly different regimes due to the effect of seasonality. 

\begin{table}[h]
	\caption{\label{tab:bgdiff}The distance in the cosine-weighted $L^2$-norm between the average geopotential height of the five months and different averaging periods.}
	\centering
	\begin{tabular}{l | ccccc }
		& November & December & January & February & March \\
		\hline
		January & 1.90 & 0.69 & 0.0 & 0.27 & 0.92 \\
		January - February & 1.89 & 0.69 & 0.13 & 0.13 & 0.85 \\
		December - February & 1.67 & 0.46 & 0.25 & 0.28 & 0.69  \\
		December - March & 1.56 & 0.38 & 0.41 & 0.36 & 0.52 \\
		November - March  & 1.24 & 0.20 & 0.69 & 0.66 & 0.39 \\
	\end{tabular}
\end{table}

We thus conclude that including November in the months for the average background state is expected to strongly bias the results, while considering only January and February is expected to be reliable. However, using only these two months means that little data is used and this lack of data might decrease the reliability of the results. Therefore, it is preferable to include more months in the analysis. If December is included in the average the difference with the included months jumps up, but when March is included as well the jump is quite a bit smaller. This brings us to the conclusion that using the months December through March leads to the most reliable results based on the balance between having sufficient data and minimizing the effect of seasonality. In the next section we discuss the effect of this choice on the regimes that are found.

\subsection{Effect of Seasonality on the Identified Regimes}
\label{sec:regimes}

Applying the $k$-means clustering approach leads to the identification of regimes for the period December through March. To study whether seasonality has an effect on the regimes that are identified we apply the clustering algorithm also to anomaly data for other periods. We start with a brief look at the inclusion of November in the period considered and subsequently discuss the effect of excluding March from the data.

\begin{figure}[h!]
	\centering
	\includegraphics[width=1.\textwidth]{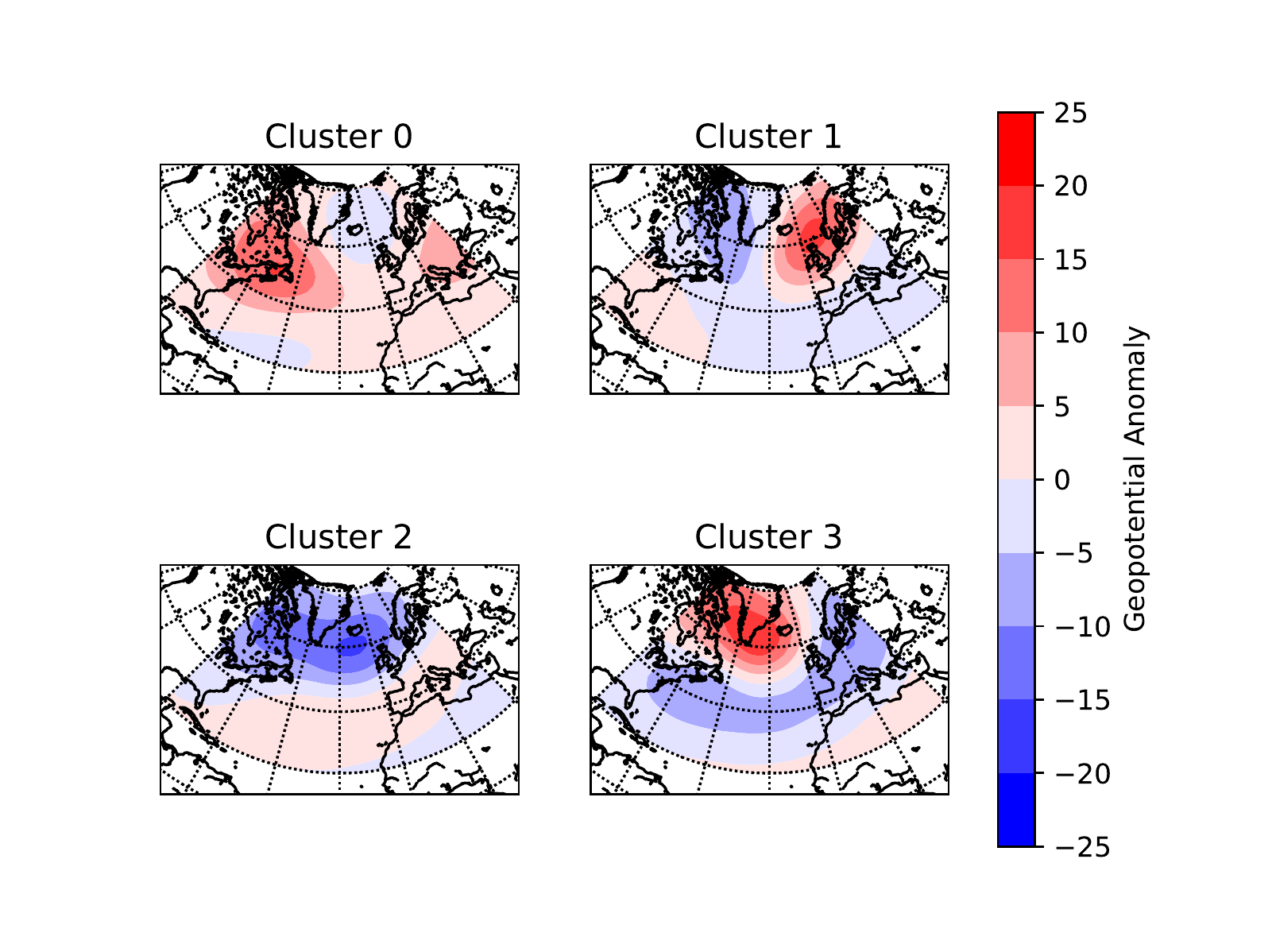}
	\vspace{-40pt}
	\caption{\label{fig:kmeans_unres_nov_k4}The clustering results of the standard $k$-means algorithm for the data including November applied to the full field for $k=4$.}
\end{figure}

When studying atmospheric circulation regimes often the period November through March is considered \citep[e.g.][]{Straus2017}. The anomalies for this period are computed with respect to a seasonally varying background state, instead of a fixed one. When a fixed background state for this period is considered the regimes found for $k=4$ are shown in Figure \ref{fig:kmeans_unres_nov_k4}. The AR is no longer found and instead a regime strongly resembling the difference in average geopotential between January and November is found, indicating that the appearance of the regime is solely due to the inclusion of November in the data. Therefore, when a fixed background state is considered, November needs to be excluded in order to obtain reliable results. The same conclusion can be drawn for $k=6$.

\begin{figure}[h!]
	\centering
	\includegraphics[width=1.\textwidth]{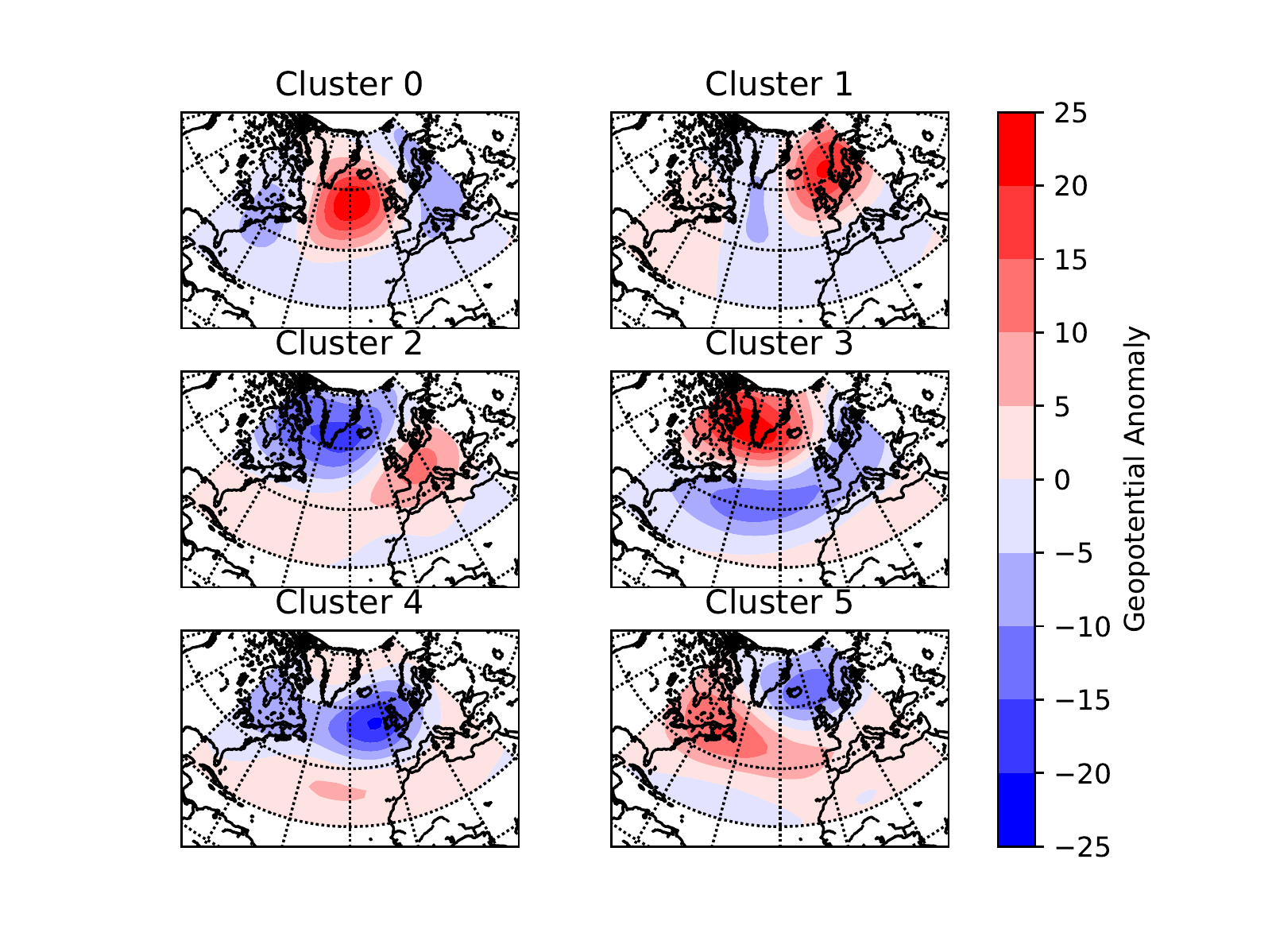}
	\vspace{-40pt}
	\caption{\label{fig:kmeans_unres_mar_k6}The clustering results of the standard $k$-means algorithm for the data without March applied to the full field for $k=6$.}
\end{figure}

For the regimes found in the paper the difference in average geopotential height between January and March bears some resemblance to a combination of the NAO- and SB- regime found for $k=6$. Therefore it is important to verify whether removing March from the period considered significantly affects the regimes that are found. The result of $k$-means clustering applied to the data for December through February is shown in Figure \ref{fig:kmeans_unres_mar_k6}. The essence of the regimes found is the same as when March is included in the data. Only small changes in the strength of the high and low pressure areas are found. The same holds for the results obtained for $k=4$.

The next question to ask is whether the occurrence and self-transition probability of the regimes is affected significantly when March is excluded from the data. For the self-transition probability the answer is No when six regimes are considered. A slight decrease of the self-transition probability of the two phases of the NAO is found for $k=4$. This decrease, even if just significant, needs to be kept in mind when considering the results but is not strong enough to provide an argument for reconsidering the period studied. The occurrence rate of the NAO+ also slightly decreases for $k=4$, whereas for the other regimes no significant changes are found in the occurrence. Finally for $k=6$ only the occurrence of the NAO- regime shows a significant change, being smaller than when March is included in the data. This means that the NAO- occurs relatively more often in March and this is the only difference that is strong enough to possibly argue for excluding March from the period considered. However, because it is only one of many aspects that show such a change we decide to stick to the period December through March for the benefit of the additional data, meanwhile keeping the difference in mind when discussing the occurrence of the regimes.

\subsection{Effect of Seasonality on Occurrence and Self-Transition Probabilities}
\label{sec:occtr}

\begin{figure}[h]
	\centering
	\begin{subfigure}{.49\textwidth}
		\centering
		\includegraphics[width=1.\textwidth]{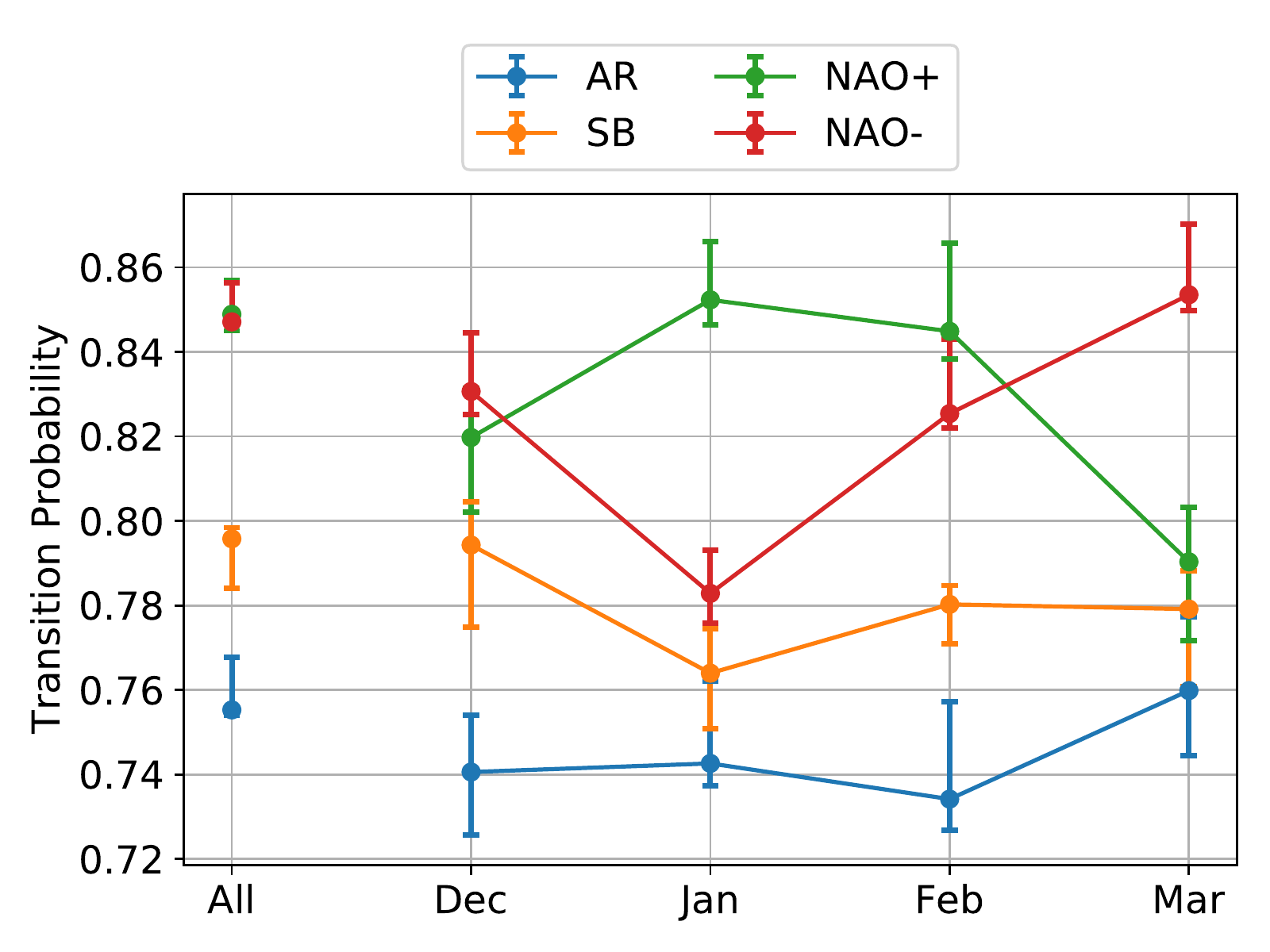}
		\vspace{-18pt}
		\caption{\label{fig:spers_C_k4}The self-transition probabilities for $k=4$.}
	\end{subfigure}
	\begin{subfigure}{.49\textwidth}
		\centering
		\includegraphics[width=1.\textwidth]{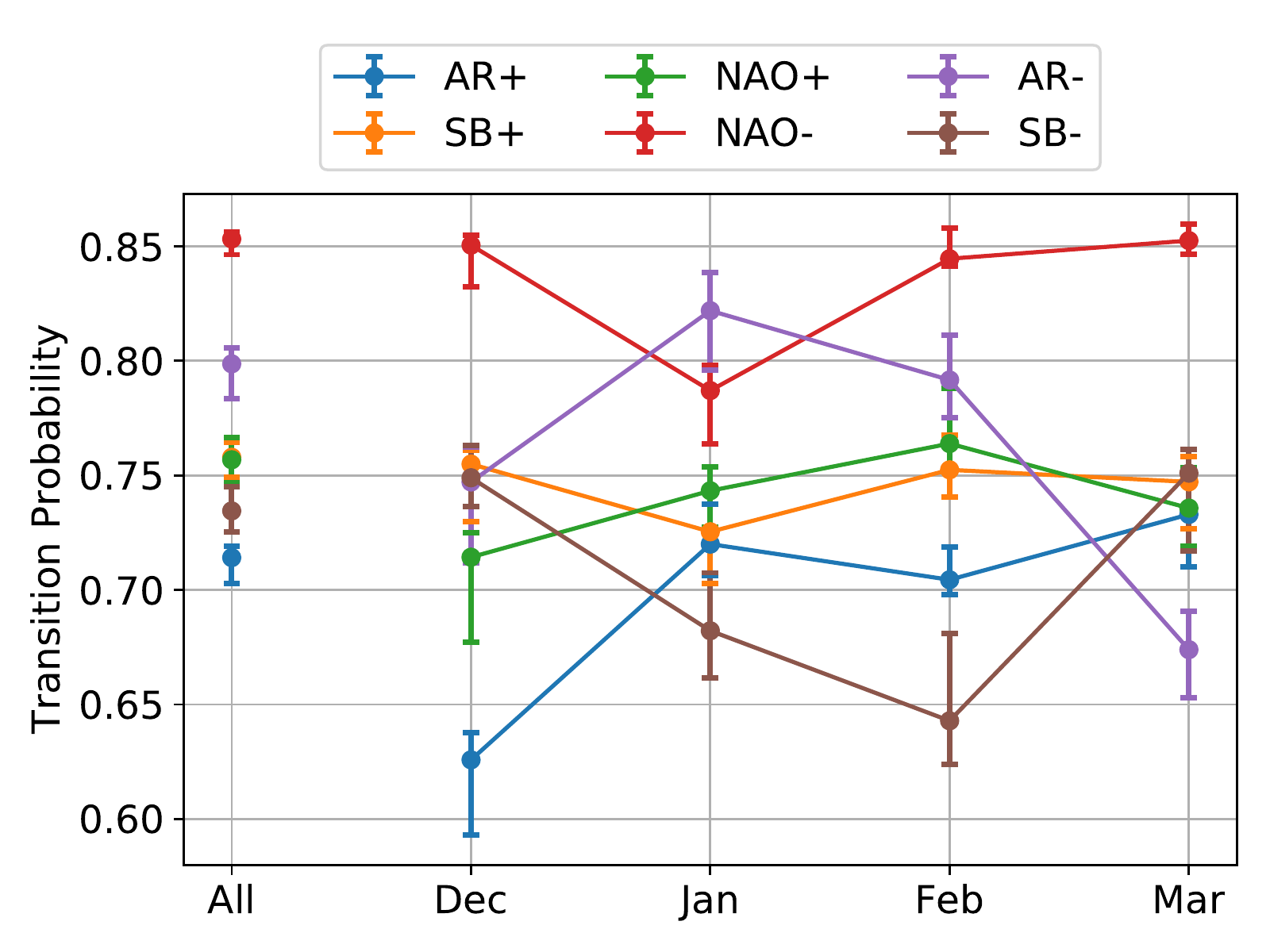}
		\vspace{-18pt}
		\caption{\label{fig:spers_C_k6}The self-transition probabilities for $k=6$.}
	\end{subfigure}
	\begin{subfigure}{.49\textwidth}
		\centering
		\includegraphics[width=1.\textwidth]{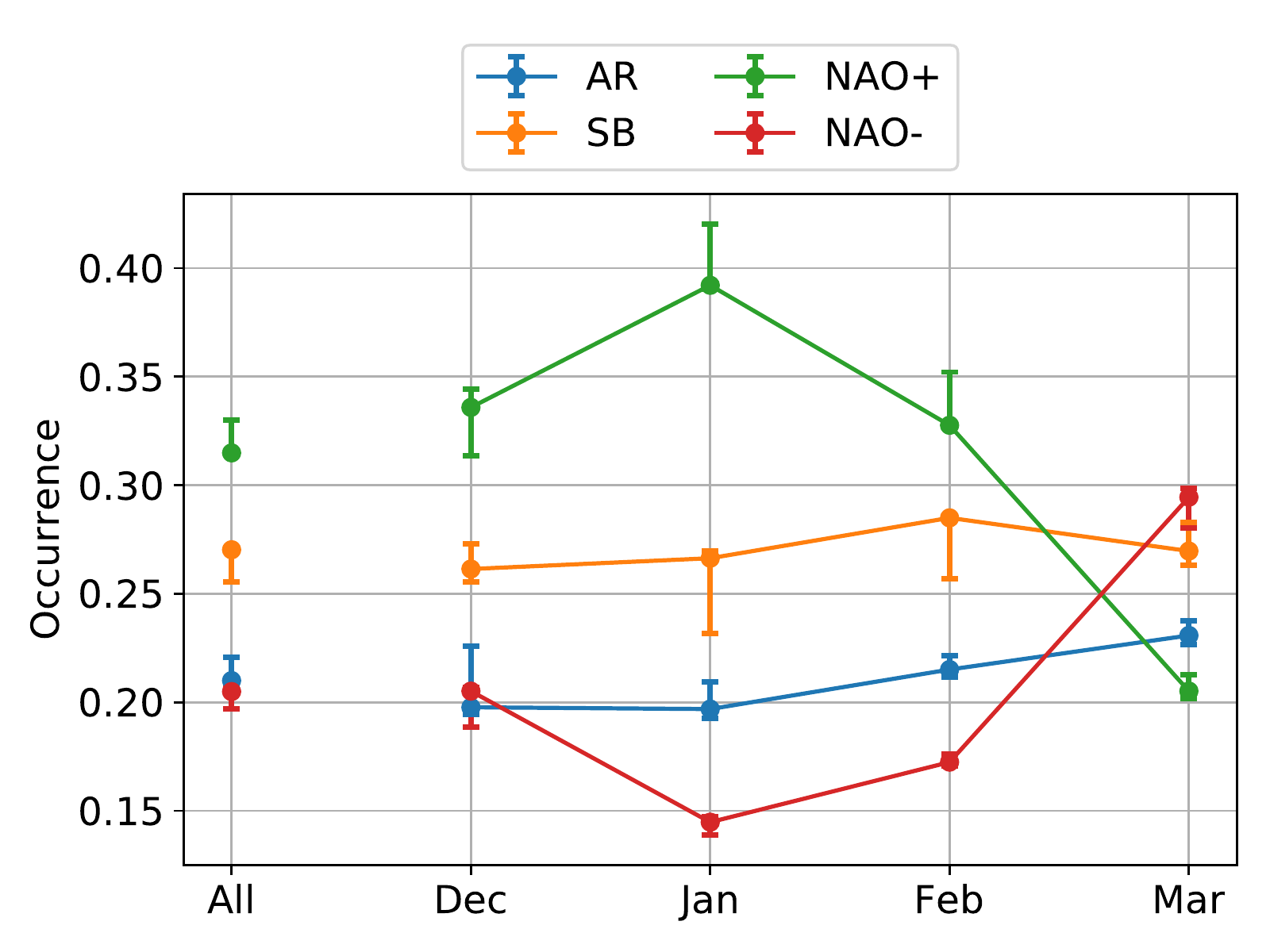}
		\vspace{-18pt}
		\caption{\label{fig:socc_C_k4}The occurrence rates for $k=4$.}
	\end{subfigure}
	\begin{subfigure}{.49\textwidth}
		\centering
		\includegraphics[width=1.\textwidth]{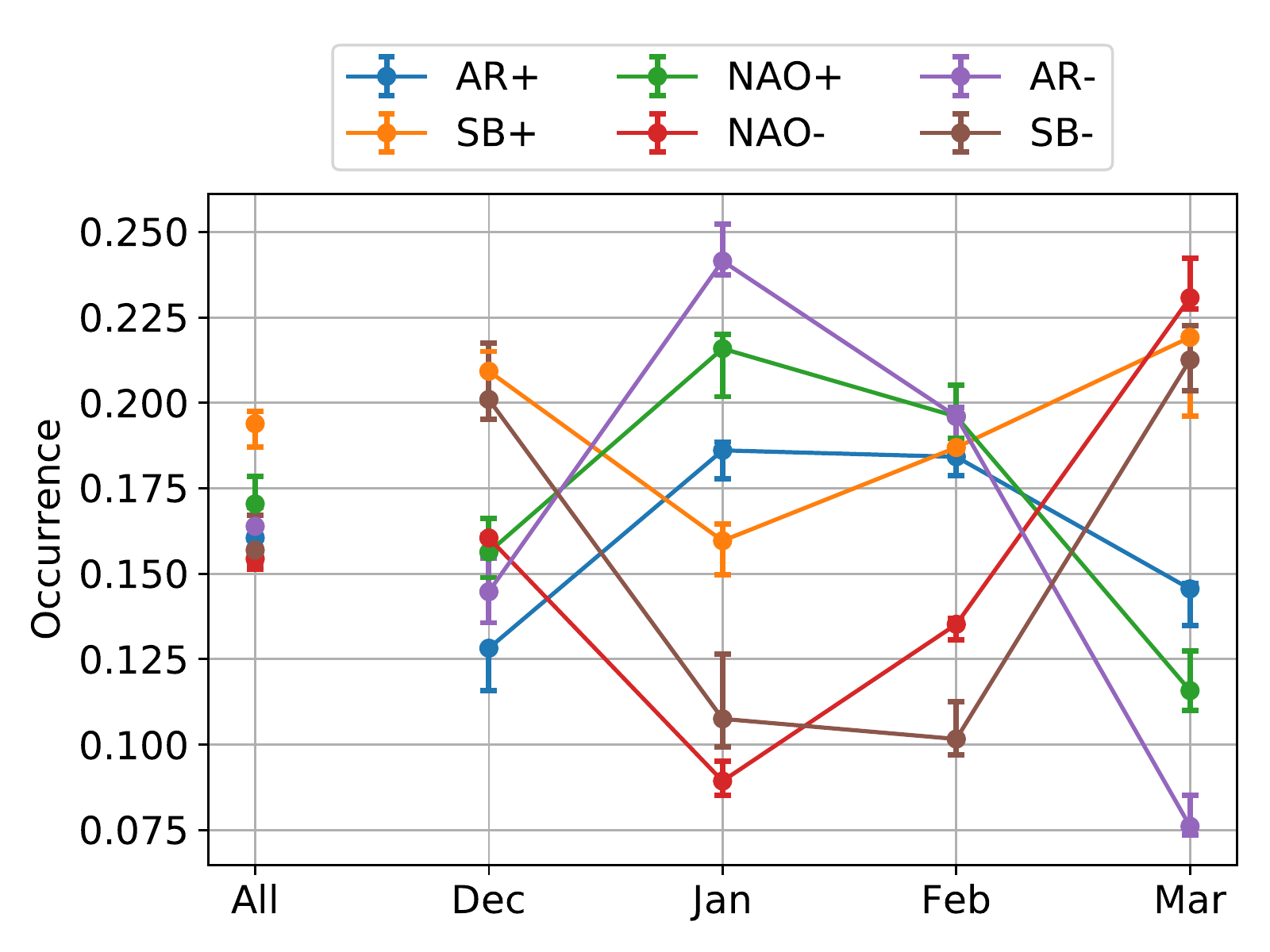}
		\vspace{-18pt}
		\caption{\label{fig:socc_C_k6}The occurrence rates for $k=6$.}
	\end{subfigure}
	\caption{\label{fig:occpers_sea}The occurrence rates and self-transition probabilities of the different regimes for $k=4$ and $k=6$ for the clustering results throughout the winter months. The error bars indicate the maximum and minimum value of occurrence/transition probability for clustering results with a slightly smaller $\mathbf{L}$ (the used bound on $\Delta\mathbf{L}$ is 0.002).}
\end{figure}

Since a constant background state (fixed climatology) is assumed instead of a seasonal varying ones, there is variability in the occurrence and transition probabilities throughout the season. As discussed in the previous sections the regimes found will not change significantly if a seasonal cycle is subtracted. However, there may be slight changes in the occurrence and persistence of some of the regimes. In Figure \ref{fig:occpers_sea} the occurrence rates and self-transition probabilities throughout the winter months are shown for both $k=4$ and $k=6$. Some regimes, like the Atlantic Ridge for $k=4$, show consistent behaviour throughout all months, while others, like the AR- for $k=6$, show significant changes in occurrence and self-transition probability throughout winter. Note that for most regimes the occurrence and self-transition probability co-vary.

\newpage

\section{Technical Details}
\label{sec:tech}

\subsection{Time-Filtering}
\label{sec:filter}

There are numerous ways in which data can be filtered to get rid of the high-frequency oscillations. Therefore we detail the method used to obtain the filtered data. The low-pass filter we use is the $\sinc$-function multiplied by the Blackman window \citep{Smith2002Book}. This window is applied because otherwise the filter never fully reaches zero. The Blackman window is given by:
\begin{equation}
\label{eq:blackman}
w(n) = 0.42 +0.5 \cos\Big( \frac{2\pi n}{N-1} \Big) + 0.08 \cos\Big( \frac{4\pi n}{N-1} \Big),
\end{equation}
where $N$ is the total number of points and $n\in[0,N-1]$. Multiplying the Blackman window with the $\sinc$ function results in the windowed $\sinc$-filter:
\begin{equation}
\label{eq:windowsinc}
h(n) = \sinc\Big( 2 f_c \Big(n - \frac{N-1}{2} \Big)\Big) \cdot w(n),
\end{equation}
where $f_c$ is the cut-off frequency. The time-filtered data is computed by convolving the filter with the time-series for each grid point.

\subsection{Toy Example Persistence Constraint}
\label{sec:toy}

The toy model which we use to study the effect of the persistence constraint consists of three clusters. The cluster centres and corresponding variances are $\mathbf{x}_1=(0.2,0.4)$, $v_1=0.012$, $\mathbf{x}_2=(0.8,0.3)$, $v_2=0.018$ and $\mathbf{x}_3=(0.7,0.9)$, $v_3=0.01$. The data is normally distributed around the cluster centre with given variance. The transition matrix used to generate the data time series is
\begin{equation}
T = \begin{pmatrix}
0.92 & 0.02 & 0.06 \\
0.04 & 0.95 & 0.01 \\
0.02 & 0.1 & 0.88 \\
\end{pmatrix}.
\end{equation}
A time series of 5000 steps is generated. For the clustering we use $k=3$ and the persistence constraint used is $C=252$.

\subsection{Information Criterion}
\label{sec:IC}

Here we briefly discuss the details of the Akaike and Bayesian information criteria (AIC and BIC). The number of parameters for the clustering is $K = k \cdot m + (k-1) \cdot T_{nr}$, $m$ dimension of the data, i.e. latitude times longitude or number of EOFs and $T_{nr}$ the number of time steps. In Table \ref{tab:IC_pt} the values of the different terms in both information criteria are shown. Here one can see that the penalty term from the BIC is too strong for the EOF data, while that of the AIC is not strong enough for the full field data.

\begin{table}[h!]
	\centering
	\caption{\label{tab:IC_pt} The values of the different terms in the AIC and BIC for both the EOF and full field data.}
	\begin{tabular}{l || ccc | ccc}
		& \multicolumn{3}{c|}{20 EOFs ($\cdot10^4$)} & \multicolumn{3}{c}{Full Field ($\cdot10^4$)} \\
		$k$ & $-2\log(\mathcal{L})$ $(26\cdot10^5 + )$ & $2K$ & $K \log(n)$ & $-2\log(\mathcal{L})$ $(17\cdot10^6 + )$ & $2K$ & $K \log(n)$ \\
		\hline
		\hline
		2 & 14.04 & 0.95 & 5.46 & 153.59 & 1.44 & 11.2 \\
		3 & 12.50 & 1.90 & 10.9 & 132.75 & 2.63 & 20.5 \\
		4 & 11.33 & 2.85 & 16.3 & 117.22 & 3.82 & 29.7 \\
		5 & 10.43 & 3.80 & 21.8 & 105.11 & 5.01 & 39.0 \\
		6 & 9.68 & 4.75 & 27.2 & 95.29 & 6.20 & 48.3 \\
		7 & 9.16 & 5.70 & 32.7 & 88.41 & 7.39 & 57.5 \\
		8 & 8.70 & 6.65 & 38.1 & 82.30 & 8.58 & 66.8 \\
		9 & 8.27 & 7.60 & 43.6 & 76.68 & 9.77 & 76.1 \\
		10 & 7.89 & 8.55 & 49.0 & 71.68 & 11.0 & 85.4 \\
	\end{tabular}
\end{table}

\subsection{Relating Average Regime Duration, Self-Transition Probability and e-Folding Time Scale}
\label{sec:efold}

In this section we focus on the derivation of relationships between the self-transition probabilities and the corresponding e-folding timescale and average regime duration. The main assumption for deriving these relations is that the regime dynamics is approximately a first-order Markov chain, meaning the current state fully determines the probability of the state the next day.

Let $p$ be the probability of a regime transitioning into itself. Consider the exponential $e^{-t/T_e}$, where $t$ is time and $T_e$ the e-folding time (both in days), describing the decay in likelihood of the atmosphere still being in the same regime after time $t$. For $t=1$ day we can relate the transition probability to the e-folding time scale by
\begin{equation}
\label{eq:Te}
\begin{split}
\e^{-1/T_e} &= p, \\
-1/T_e &= \log(p), \\
T_e &= -\frac{1}{\log(p)},
\end{split}
\end{equation}
meaning that if we know one, we can compute the other.

The next step is to relate the self-transition probability to the expected (or average) regime duration. Starting from a regime with self-transition probability $p$ (day 0), the expected time it takes to transfer out of that regime ($O$) is
\begin{equation}
\label{eq:Tav}
\begin{split}
\mathbb{E}(\text{days to } O) &= \sum_{n=1}^{\infty} n P(O \text{ at day } n), \\
&= \sum_{n=1}^{\infty} n p^{n-1} (1-p), \\
&= (1-p) \sum_{n=0}^{\infty}\frac{\d}{\d p} p^{n}, \\
&= (1-p) \frac{\d}{\d p} \sum_{n=0}^{\infty} p^{n}, \\
&= (1-p) \frac{\d}{\d p} \Big(\frac{1}{1-p} \Big), \\
&= (1-p) \frac{1}{(1-p)^2} = \frac{1}{1-p}.
\end{split}
\end{equation}
We denote the found average regime duration by $\mathbb{E}(\text{days to } O) = T_{av}$. Note that the Taylor series of $-\log(p)$ around 1 is
\begin{equation}
-\log(p) = (1-p) + \mathcal{O}(p^2),
\end{equation}
meaning that in the limit $p\rightarrow1$ we have that $T_e$ and $T_{av}$ become equal (by linking Equation \eqref{eq:Te} and \eqref{eq:Tav}).

The found relations between $p$ and either $T_e$ or $T_{av}$ allow us to also express the e-folding time $T_e$ as a function of $T_{av}$ and the other way around. This yields
\begin{equation}
T_e = -\frac{1}{\log(T_{av}-1) - \log(T_{av})}, \qquad T_{av} = \frac{1}{1 - \exp(-1/T_e)}.
\end{equation}

\section{Consistency Results}
\label{sec:consist}

Figures showing the distributions of $\Delta\mathbf{L}$ and the data similarity are shown for different datasets and clustering algorithms.

\begin{figure}[h]
	\centering
	\begin{subfigure}{.95\textwidth}
		\centering
		\includegraphics[width=1.\textwidth]{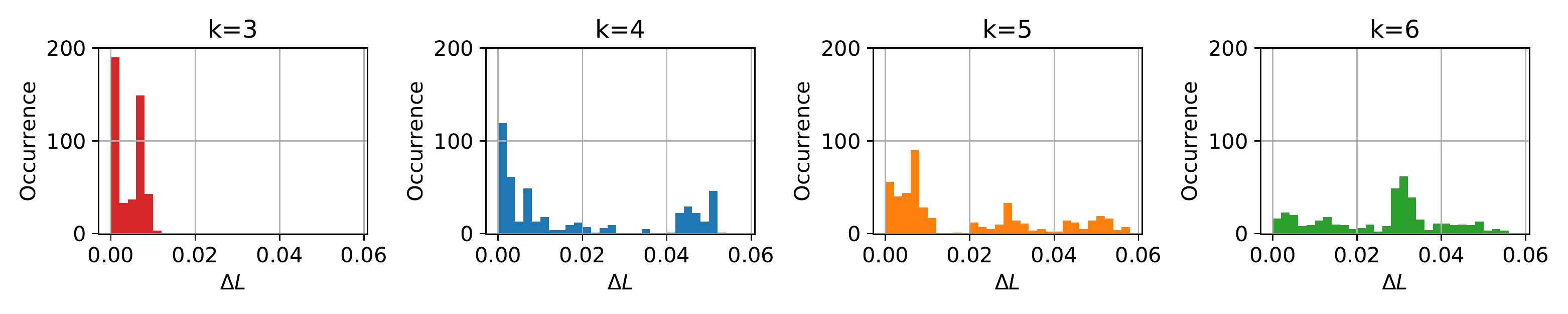}
		\vspace{-20pt}
		\caption{\label{fig:L_k47_oe1}The clustering functional $\mathbf{L}$.}
	\end{subfigure}
	\begin{subfigure}{.95\textwidth}
		\centering
		\includegraphics[width=1.\textwidth]{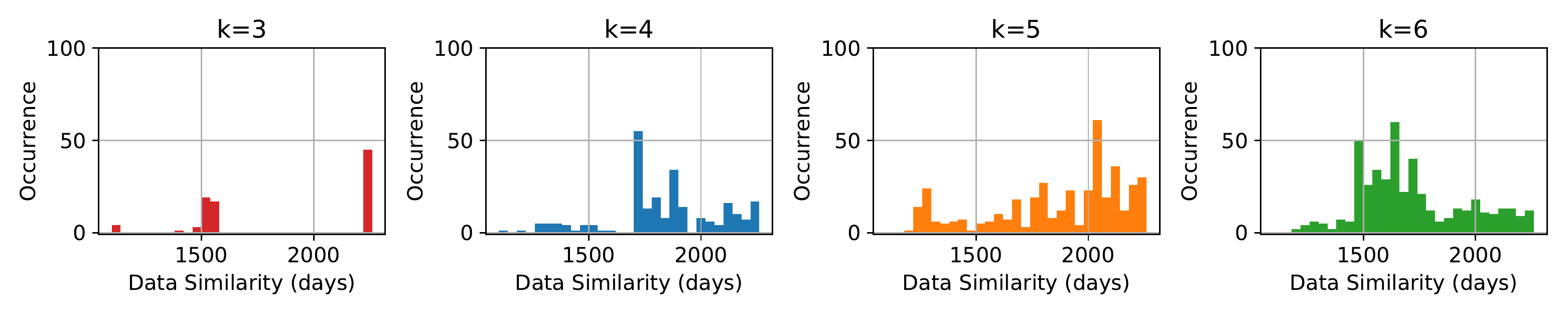}
		\vspace{-20pt}
		\caption{\label{fig:Data_k47_oe1}The data similarity.}
	\end{subfigure}
	\caption{\label{fig:sim_k47_oe1}Histograms for the clustering functional $\mathbf{L}$ and the data similarity with respect to the optimal (minimal $\mathbf{L}$) result using the odd years of the full field data for $k=3,...,6$.}
\end{figure}

\begin{figure}[h]
	\centering
	\begin{subfigure}{.95\textwidth}
		\centering
		\includegraphics[width=1.\textwidth]{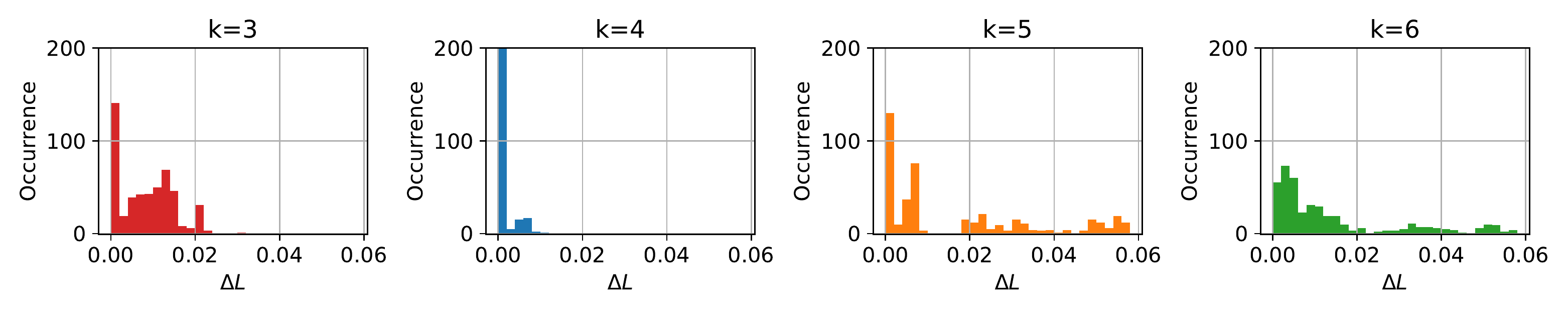}
		\vspace{-20pt}
		\caption{\label{fig:L_k47_oe2}The clustering functional $\mathbf{L}$.}
	\end{subfigure}
	\begin{subfigure}{.95\textwidth}
		\centering
		\includegraphics[width=1.\textwidth]{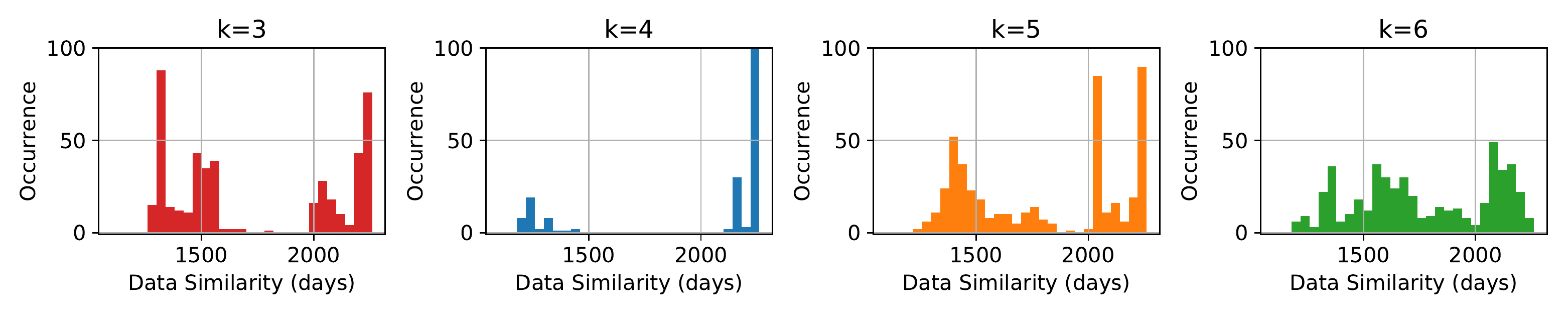}
		\vspace{-20pt}
		\caption{\label{fig:Data_k47_oe2}The data similarity.}
	\end{subfigure}
	\caption{\label{fig:sim_k47_oe2}Histograms for the clustering functional $\mathbf{L}$ and the data similarity with respect to the optimal (minimal $\mathbf{L}$) result using the even years of the full field data for $k=3,...,6$.}
\end{figure}

\section{The Lorenz 63 System}
\label{sec:system}

The Lorenz 63 system \citep{Lorenz1963} is one of the most well-studied low-order systems derived from geophysical fluid dynamics. It exhibits regime behaviour, as well as chaos, making it a suitable system to test the performance of clustering methods \citep[e.g.][]{Hannachi2001}, but also to use as an analogue for more complex systems in climate. For example \citet{Corti1999} used this feature to interpret the effect of climate change on atmospheric circulation regimes. Here we apply the same clustering approaches as used to identify atmospheric circulation regimes in the main article to different realisations of the Lorenz 63 system. This allows for testing the accuracy and reliability of the methods used.

The equations of the Lorenz 63 system are \citep{Lorenz1963}:
\begin{equation}
\begin{split}
\frac{\d x}{\d t} &= -\sigma x + \sigma y, \\
\frac{\d y}{\d t} &= -x z + r x - y, \\
\frac{\d z}{\d t} &= x y - b z.
\end{split}
\end{equation}
Here $\sigma$, $r$ and $b$ are parameters. The standard values used by Lorenz are $\sigma = 10$, $r=28$ and $b = 8/3$ and give the well-known butterfly. We stick to these values here. The system is integrated using a standard Euler scheme with time steps of $10^{-2}$ for $10^4$ steps.

\subsection{$k$-means Clustering}
\label{sec:kmeans}

\begin{figure}[h]
	\centering
	\begin{subfigure}{.49\textwidth}
		\centering
		\includegraphics[width=1.\textwidth]{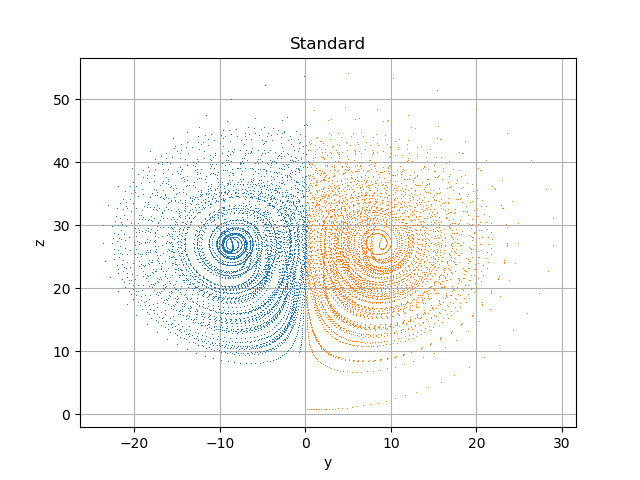}
		\caption{\label{fig:l63_yz}$y$-$z$-plane.}
	\end{subfigure}
	\begin{subfigure}{.49\textwidth}
		\centering
		\includegraphics[width=1.\textwidth]{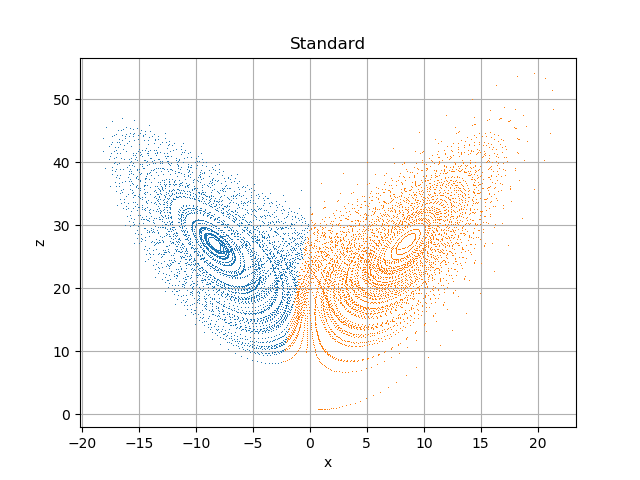}
		\caption{\label{fig:l63_xz}$x$-$z$-plane.}
	\end{subfigure}
	\caption{\label{fig:l63}Clustering of the Lorenz 63 system in either the $y$-$z$- or $x$-$z$-plane. The dots show the data points of the simulation of the model and the colours indicate to which cluster these points are assigned.}
\end{figure}

We apply the $k$-means clustering algorithm to several realisations of the Lorenz 63 system (different initial conditions) projected onto either the $y$-$z$-plane or the $x$-$z$-plane. Mostly the clusters found in the $y$-$z$-plane correspond to the `correct' clusters, being the two wings of the butterfly as can be seen in Figure \ref{fig:l63_yz} separated by the line $y=0$. On the other hand, when the $k$-means algorithm is applied to the corresponding $x$-$z$-data the result is not as good since it fails in identifying the two wings of the butterfly as the clusters, separated by the $x=0$ line (Figure \ref{fig:l63_xz}). The standard $k$-means algorithm thus fails to correctly identify the two wings of the butterfly in the Lorenz 63 system when only $x$ and $z$ data is taken into account.

To improve the result of the $k$-means algorithm for the $x$-$z$-plane we explore two methods that enforce consistency in time of the clustering result. The first method is to apply a low-pass filter to the data getting rid of high-frequency oscillations. The second method is to include a persistence constraint in the $k$-means algorithm itself. For both approaches a parameter has to be chosen. For the time-filter this is the cut-off frequency and for the persistent algorithm this is the value of the persistence constraint.

The approach taken here for choosing these two parameters is mirrored to the approach in the main article. This means we a priori set the cut-off frequency for the time-filtering to 150 time steps, being the equivalent of the 10-day filter used for the atmospheric circulation regimes. On the other hand the value of the persistence constraint is determined using an information criterion. Both the Akaike Information Criterion (AIC) and Bayesian Information Criterion (BIC) are explored and in the end the BIC is chosen as the better criterion. This is because when the two criteria differ in the location of their minimum the clusters for the BIC-minimum are closer to the best result than those for the AIC-minimum. Note that the information criterion approach can also be taken to determine the optimal cut-off frequency for the time-filtering.

\begin{figure}[h]
	\centering
	\begin{subfigure}{.49\textwidth}
		\centering
		\includegraphics[width=1.\textwidth]{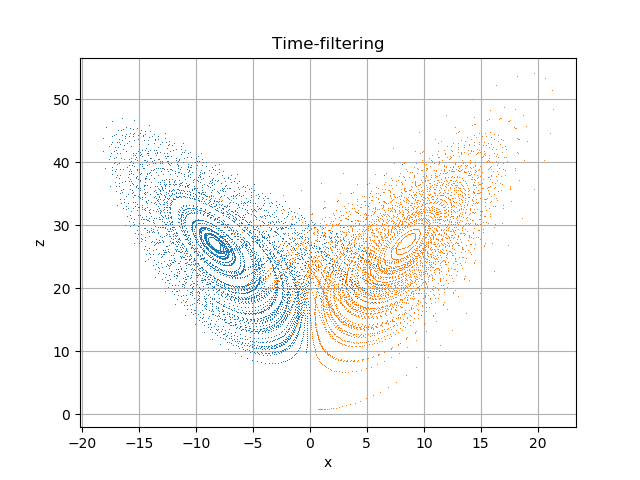}
		\caption{\label{fig:l63_time}Applying a low-pass filter to the data for a cut-off frequency of 150 time steps.}
	\end{subfigure}
	\begin{subfigure}{.49\textwidth}
		\centering
		\includegraphics[width=1.\textwidth]{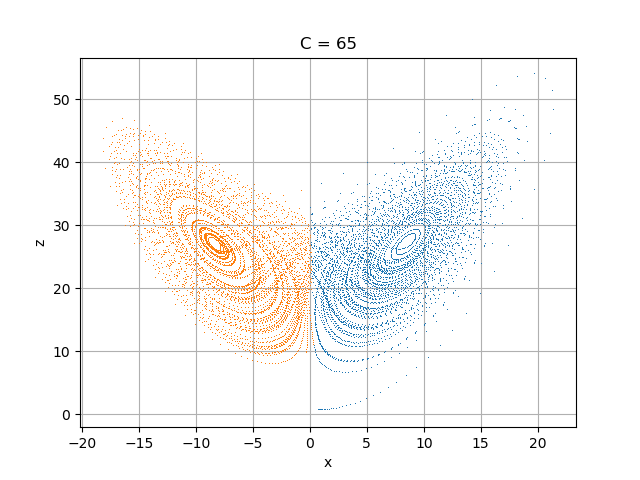}
		\caption{\label{fig:l63_pers}Incorporating a persistence constraint in the clustering algorithm.}
	\end{subfigure}
	\caption{\label{fig:l63_filter}Clustering of the Lorenz 63 system in the $x$-$z$-plane using different methods to enforce persistence.}
\end{figure}

For the same realisation as in Figure \ref{fig:l63} the results for the time-filtered data and clustering using a persistence constraint are shown in Figure \ref{fig:l63_filter}. Both results show a clear improvement with respect to the standard approach in Figure \ref{fig:l63_xz}. The result for the time-filtered data shows a slightly different assignment of data to clusters for the transition trajectories, but other than that the results both are as desired. We note that for this realisation of the model the BIC shows a clearly identifiable minimum, which is not always the case.

\begin{figure}
	\centering
	\begin{subfigure}{.49\textwidth}
		\centering
		\includegraphics[width=1.\textwidth]{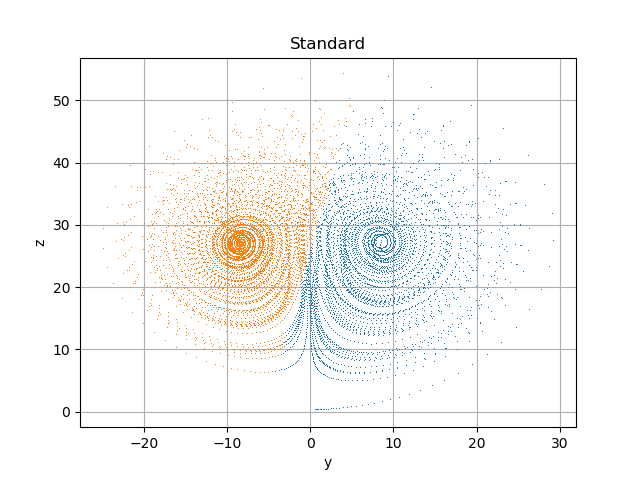}
		\caption{\label{fig:l63_yz2}Standard $k$-means applied to the $y$-$z$-data.}
	\end{subfigure}
	\begin{subfigure}{.49\textwidth}
		\centering
		\includegraphics[width=1.\textwidth]{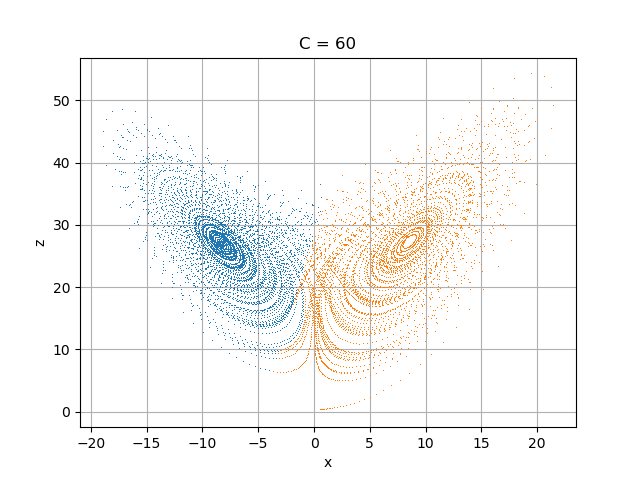}
		\caption{\label{fig:l63_pers2}Incorporating a persistence constraint in the clustering algorithm.}
	\end{subfigure}
	\caption{\label{fig:l63_off}The clustering approaches discussed do not always identify the correct clusters.}
\end{figure}

The discussed results show a realisation of the model in which both methods work nicely. This however is not the case for every realisation of the Lorenz 63 model. Already when applying $k$-means clustering to the $y$-$z$ data the correct clusters are not always identified, as can be seen in Figure \ref{fig:l63_yz2}. This result likely improves if more data is included, but as a limited amount of data is one of the difficulties of real world clustering it is important to note this limitation. Furthermore the BIC does not always point towards the correct result, which can be seen in Figure \ref{fig:l63_pers2}. By looking at the clusters for different values of the constraint it is possible to identify a better value, but as this is impossible for the high-dimensional atmospheric data in which the circulation regimes are identified this is not a desirable option. We note that also for the time-filtered data the result is not always as good as shown in Figure \ref{fig:l63_time}, although in general it is slightly more robust than the results for the persistence constraint.

Applying the clustering methods used to identify circulation regimes in atmospheric data to the Lorenz 63 system teaches us to be careful in relying too much on the outcome of the algorithm. Even for the `simple' Lorenz 63 system the clustering algorithm does not always identify the correct clusters. For the even more complex atmospheric data in which the circulation regimes are identified this is likely an even larger difficulty. This does not mean the result is not useful, but it is important to be aware of the limitations of the method.

\end{document}